\documentclass[a4paper,12pt]{article}

\usepackage{graphicx}
\usepackage{dcolumn}
\usepackage{bm}
\usepackage{gensymb}
\usepackage{amsmath}
\usepackage{float}
\usepackage{authblk} 
\usepackage{abstract}

\title{Fluctuation-dissipation theorems and the measurement of the Onsager coefficients for two-phase flow in porous media}

\author[1]{Marcel Moura\thanks{Corresponding author: marcel.moura@fys.uio.no}}
\author[2]{Dick Bedeaux}
\author[3]{Ryan T. Armstrong}
\author[2]{Signe Kjelstrup}

\affil[1]{PoreLab, the Njord Center, Department of Physics, University of Oslo, Oslo, Norway}
\affil[2]{PoreLab, Department of Chemistry, Norwegian University of Science and Technology, Trondheim, Norway}
\affil[3]{School of Civil and Environmental Engineering, The University of New South Wales, Sydney, Australia}

\date{\today}

\begin{document}

\maketitle

\begin{abstract}
We propose a new methodology for the experimental measurement of the Onsager coefficients of porous media flows by application of the fluctuation-dissipation theorem. The experimental setup consists of a steady-state flow condition in which two incompressible fluids are simultaneously injected into a modified Hele-Shaw cell. The cell is transparent and allows direct visualization of the dynamics via regular optical imaging methods. The fluctuations in the phase saturations are obtained and, by computing the temporal correlations of their time derivatives, we gain access to the Onsager coefficients. This work gives experimental grounding to recent theoretical development on the applications of the fluctuation-dissipation theorems to porous media flows.
\end{abstract}


\section{Introduction}

The fluctuation-dissipation theorems (FDT) were given their firm basis by the work of Callen and Welton \cite{Callen1951}, Green \cite{Green1954} and Kubo \cite{Kubo1966} in the 1950s and 1960s. They are well known from their frequent use in homogeneous systems to find diffusion coefficients \cite{Liu2011} or thermal conductivities \cite{Bresme2014}.  Porous media form an exception in this context. A theoretical analysis of fluctuations in porous media was given by Bedeaux and Kjelstrup \cite{bedeaux2021} followed by an experimental study by Alfazazi et al. \cite{Alfazazi2024}. 

The FDT was originally formulated for systems in equilibrium \cite{Callen1951,Green1954,Kubo1966} and is therefore most commonly used in such systems. The fluctuations leading to the Einstein relation can be taken as a central example for the determination of diffusion coefficients (zero frequency time correlations). The FDT can also be formulated for fluctuations of dissipative fluxes in driven systems. This was discussed in detail in the book by Ortiz de Zarate and Sengers \cite{Ortiz2006}. The FDT remains valid for these fluxes, but long-range hydrodynamic fluctuations of, for instance, the fluid density appear when a temperature gradient is applied (static space correlations). Experimental evidence was presented \cite{Ortiz2006}. 

In their recent work, Bedeaux and Kjelstrup \cite{bedeaux2021} gave theoretical expressions for the time and space correlations of the volume fluxes of fluids in a porous medium. Winkler et al. \cite{winkler2020} made a first network simulation to determine these correlations. A symmetry in the correlations was established, but a link was not made to either Onsager coefficients or experimental work. The first such link for zero frequency time correlation functions was made by Alfazazi \textit{et al.} \cite{Alfazazi2024}. The present work aims to develop another experimental procedure for the study of velocity fluctuations for two-phase flow in a Hele-Shaw cell. We shall derive expressions and report experiments on zero frequency time correlations in the system and relate the correlations to the Onsager coefficients.  

The assumption of local equilibrium in the representative elementary volume (REV) of a porous medium plays a central role in the development of a continuum theory for the related transport phenomena \cite{Kjelstrup2018,Kjelstrup2019}. The FDT is here formulated in terms of the fluctuating fluxes \cite{bedeaux2021,Bedeaux2024} that apply to a REV in local equilibrium. The aim of the present investigation is to find the conditions that enable us to apply FDT for porous media flows. Velocity fluctuations in multiphase flows in porous media can encode information about the medium's hydraulic permeability. Porous media permeabilities are extensively used for description of flows on the continuum level and are traditionally found from a plot of the steady state volumetric flow rate versus an effective pressure gradient \cite{bear1972}. Alternative methods are of interest \cite{Alfazazi2024}.

Flow in porous media typically involves a series of general phenomena giving rise to fluctuations, such as ganglion dynamics \cite{tallakstad2009,avraam1995,rucker2015,sinha2017,eriksen2022} Haines jumps \cite{haines1930,furuberg1996,maloy1992,moura2017,moura2017_2,spurin2022} and intermittent avalanche-like behavior \cite{moura2020,maloy2021,spurin2021,armstrong2014,heijkoop2024}. These fluctuations can either be in the pressure signal, phase velocities or pore configurations and lead to the characteristic flow front developments observed \cite{rothman1990,grunde2004,moura2020,zhao2016,bauyrzhan2018,brodin2022}. It has also been shown that two-phase flows reach a history-independent steady-state \cite{erpelding2013} in the sense that macroscopic properties such as the pressure difference across the model and the saturation reach a stationary state, oscillating around well defined mean values. 
There is also growing evidence that the noise in system variables (i.e. in the pressure) is characteristic for the type of flow, see \cite{Alfazazi2024} for an overview. We expect the time scales of the correlations to be important, also for fluctuations in extensive variables as those studied here.   

In this paper we outline an experimental technique to measure the Onsager coefficients for two-phase porous media flows via optical measurements of fluctuations in phase saturation. Unlike the method used by Alfazazi et al. \cite{Alfazazi2024}, volumetric fluxes are not measured. We study instead the statistics of the time derivative of the phase saturation and show that under steady-state conditions, this quantity fluctuates around zero (as expected) and follows a Gaussian distribution (at least for small fluctuations). We compute the autocorrelation function of the fluctuations and its integral, which then gives access to the Onsager coefficients for the case of two incompressible fluid phases. We demonstrate two alternative methods of arriving at this quantity, one considering a time average of the whole experimental signal, and another in which we produce a pseudo-ensemble of subexperiments by clipping the time-signal and averaging the results computed for each sub-experiment. The results of both techniques should be identical, but offer different perspectives to the problem: while the full time-averaging method is more straightforward and easier to compute, the pseudo-ensemble averaging procedure gives a better understanding of the statistical convergence of the measures and the characteristics of the underlying phenomena.

The paper is divided as follows: in Section \ref{sec:description} we present the experimental setup employed and boundary conditions driving the flows. Section \ref{sec:th_fr} is dedicated to the theoretical framework with the application of the fluctuation-dissipation theorem for porous medium flows. In particular, we show how correlations in saturation can be used to extract the Onsager coefficients. Section \ref{sec:results} shows our experimental results and describes the different approaches to the measurements. Finally in Section \ref{sec:conclusions} we present the conclusions and discuss possible extensions of the work.

\section{Description of the experimental setup}
\label{sec:description}

\begin{figure}[t]
	\centering
	\includegraphics[width=0.5\linewidth]{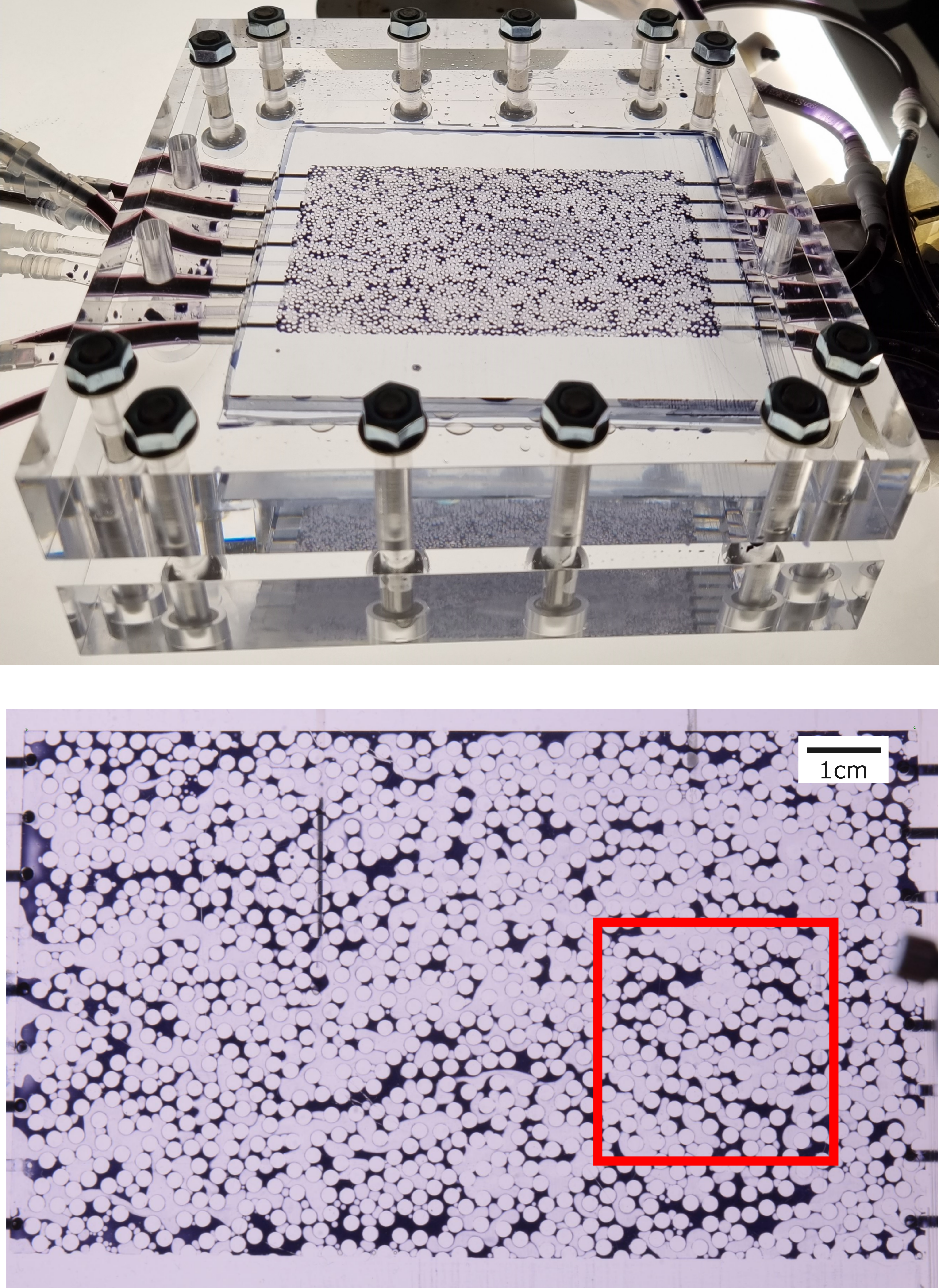}
	\caption{Experimental setup consisting of a modified Hele-Shaw cell with a 3D printed porous network. The flow direction is from left to right and the tubes on the left side of the model are connected to two syringe pumps. On the bottom image, we see a snapshot of the process. The red square denotes the approximate field of view of the machine vision camera used for the imaging at higher frame rate.}
	\label{fig:syst}
\end{figure}

We have employed a custom-built synthetic transparent network to study the dynamics of a steady-state experiments in porous media, see Fig.~\ref{fig:syst}. A description of a similar setup can also be found in Refs.~\cite{maloy2021,vincentdospital2022}. The setup consists of a modified Hele-Shaw cell \cite{heleshaw1898}, formed by a porous structure placed between 2 plexiglass plates. The plates are clamped around the perimeter using screws to give robustness to the setup and ensure the quasi-2D geometry. The porous medium is produced via a stereolithographic 3D printing technique (Formlabs Form 3 printer) using a transparent printing material (Clear Resin FLGPCL04). This technique allows us to control the geometry of the porous medium and to visualize the fluid dynamics via regular optical imaging. The porous model consists of a monolayer of cylinders with diameter $2$mm and height $0.5$mm that are distributed according to a Random Sequential Adsorption (RSA) algorithm \cite{hinrichsen1986}. The minimal distance between the cylinders is chosen to be $0.3$mm. The porous structure has porosity $\phi = 0.57$ and dimensions 12 cm x 7 cm, where the first number denotes the length of the network (measured in the inlet--outlet direction) and the second is its width (measured in the perpendicular direction). 9 inlet ports are placed along one side of the model and 8 outlet ports are placed on the other side. The 9 inlet ports feed the fluids into the system. Dyed water (dynamic viscosity $\eta_w = 8.9 10^{-4} Pa.s$, density $\rho_w= 997 kg.m^{-3}$) and light mineral oil (dynamic viscosity $\eta_o = 2.3 10^{-2} Pa.s$, density $\rho_o = 833 kg.m^{-3}$) were used as the fluid phases which are distributed in an alternating fashion along the inlets, with 5 inlets for water and 4 for oil. The water phase is dyed dark blue by adding a pigment to it (Nigrosin). The odd number of inlet ports ensures a top-bottom symmetry of the boundary condition of the system, although this is not a strict requirement for the experiment. Each inlet is connected to a syringe with the respective fluid. The 5 water syringes and 4 oil syringes are then controlled by 2 separate pumps (Harvard Apparatus PhD Ultra), so that the total flow rate of each fluid phase can be set independently. 

A high-resolution camera (Nikon D7100) records the experiment with pictures every 30 seconds. Preliminary experiments showed that the typical decorrelation time for the dynamics was too fast to be captured with this frame rate, so we opted to employ a secondary machine vision camera (Allied Vision Alvium 1800 U-158c) for filming a smaller segment of the model at a higher frame rate (10 frames per second). The approximate field of view of this second camera is shown on the bottom part of Fig.~\ref{fig:syst}, on top of an image of the whole experiment taken by the first camera. The core of the analysis in this paper is focused on the data from the machine vision camera, but having access to the high-resolution images of the whole model was also important. It allowed us for instance to guarantee that boundary effects from inlet (channeling due to point injection of the fluids) would dissipate before reaching the field of view of the machine vision camera \cite{erpelding2013,aursjo2014}. The boundary effects are rate- and fluid-dependent and we noticed that the higher the injection rate, the deeper the inlet channeling effects penetrate into the model, thus rendering a smaller segment of the model under a steady-state condition.

\section{Theoretical framework}
\label{sec:th_fr}
\subsection{Saturations in a Hele-Shaw cell}

In this work we have treated the Hele-Shaw cell as purely 2 -dimensional. In reality it has in fact a finite thickness in the $z$-direction. In the following discussion, when we use a quantity as a function of $x, y$ and not of $z$, this implies that we use the average of this quantity over the thickness of the cell.

For each $A=L_{x} \times L_{y}$ surface element one can at each moment in time measure the areas covered by the more wetting, $A_{w}$, and the less wetting, $A_{n}$, fluid phases, identified by the subscripts $n$ and $w$, respectively, as well as of the solid (``rock'') phase, $A_{r}$. The surface area of the pores is $A_{p}=A_{w}+A_{n}$. The porosity is defined by $\phi \equiv A_{p} / A=A_{p} /\left(A_{p}+A_{r}\right)$. We assume that $A_{p}$ and $A_{r}$ and therefore $\phi$ are time independent. The saturations of the more and the less wetting fluid phases are defined by $S_{w} \equiv A_{w} / A_{p}$ and $S_{n} \equiv A_{n} / A_{p}$. From these definitions it follows that $S_{w}+S_{n}=1$, where it is important to note that this is also the case if one or two of the fluid phases are compressible.

In the formulation of the nonequilibrium thermodynamics of porous media we used the so-called representative elementary volume (REV) \cite{Kjelstrup2018,Kjelstrup2019}. For the Hele-Shaw cell it would be more appropriate to call it the representative elementary surface area. This implies that we use a surface area $A^{\mathrm{REV}}=L_{x}^{\mathrm{REV}} \times L_{y}^{\mathrm{REV}}$, which is large enough to cover a representative selection of pores, but not larger than that. For each point $x, y$ in the cell the representative elementary surface area is defined as the surface area for which $x<x^{\prime}<x+L_{x}^{\mathrm{REV}}$ and $y<y^{\prime}<y+L_{y}^{\mathrm{REV}}$. The introduction of the REV makes it possible to give a continuous course-grained description of the porous medium in terms of averages over the REV \cite{Kjelstrup2018,Kjelstrup2019}. The fluctuation-dissipation theorems could then be derived \cite{bedeaux2021} in terms of such a coarse-grained desctiption.

On the pore level we introduce the characteristic functions $s_{w}(x, y, t), s_{n}(x, y, t)$ and $s_{r}(x, y, t)$. They are equal to 1 in the corresponding phase and zero in the other phases. Their sum is by definition 1 everywhere. The surface areas of the three phases in the REV, are given in terms of the characteristic functions by

\begin{equation}
A_{j}^{\mathrm{REV}}(x, y, t)=\int_{x}^{x+L_{x}} d x^{\prime} \int_{y}^{y+L_{y}} d y^{\prime} s_{j}\left(x^{\prime}, y^{\prime}, t\right)
\end{equation}

We have $\phi^{\mathrm{REV}} \equiv A_{p}^{\mathrm{REV}} / A^{\mathrm{REV}}=A_{p}^{\mathrm{REV}} /\left(A_{p}^{\mathrm{REV}}+A_{r}^{\mathrm{REV}}\right)$. We will assume that $A_{p}^{\mathrm{REV}}$ and $A_{r}^{\mathrm{REV}}$ and therefore $\phi^{\mathrm{REV}}$ are time independent. The saturations of the more and the less wetting fluid phases are defined by $S_{w}^{\mathrm{REV}} \equiv A_{w}^{\mathrm{REV}} / A_{p}^{\mathrm{REV}}$ and $S_{n}^{\mathrm{REV}} \equiv A_{n}^{\mathrm{REV}} / A_{p}^{\mathrm{REV}}$. From these definitions it follows that $S_{w}^{\mathrm{REV}}+S_{n}^{\mathrm{REV}}=1$. Both $S_{w}^{\mathrm{REV}}$ and $S_{n}^{\mathrm{REV}}$ can depend on $(x, y, t)$.

At each point $x, y$ along the cell the representative elementary surface area (REV) defines the value of the various quantities as averages over the REV. When one measures saturations one usually considers larger surface areas. For a surface area $A_{xy}=L_{x} \times L_{y}$, where $L_{x}>L_{x}^{\mathrm{REV}}$ and $L_{y}>L_{y}^{\mathrm{REV}}$, one has

\begin{equation}
S_{j}(x, y, t)=\frac{1}{A_{xy}} \int_{x}^{x+L_{x}} d x^{\prime} \int_{y}^{y+L_{y}} d y^{\prime} S_{j}^{\mathrm{REV}}\left(x^{\prime}, y^{\prime}, t\right)
\label{eq:SjSrev}
\end{equation}
for $j=w, n$. Similar equalities are valid for all the thermodynamic properties of the larger surface areas.

After a steady-state situation is achieved, the saturations will fluctuate around their average value. These averages can be found using

\begin{equation}
\left\langle S_{j}(x, y)\right\rangle=\frac{1}{\tau} \int_{t}^{t+\tau} d t^{\prime} S_{j}\left(x, y, t^{\prime}\right)
\end{equation}
where $\tau$ is sufficiently large. The averages of the REV variables are defined in the same way.

The quantity we are interested in, is the correlation function of the time derivative of the saturations:
\begin{equation}
\begin{aligned}
C_{ij}(\Delta x, \Delta y, \Delta t) &\equiv \left\langle \frac{\partial S_{i}(x, y, t)}{\partial t} \frac{\partial S_{j}(x+\Delta x, y+\Delta y, t+\Delta t)}{\partial t} \right\rangle \\
&= \frac{1}{\tau} \int_{t}^{t+\tau} dt' \frac{\partial S_{i}(x, y, t')}{\partial t'} \frac{\partial S_{j}(x+\Delta x, y+\Delta y, t'+\Delta t)}{\partial t'}
\label{eq:Cij}
\end{aligned}
\end{equation}

We focus now on the part of the Hele-Shaw cell that is homogeneous so that this time average only depends on $\Delta x, \Delta y, \Delta t$. This is away from the inlet along the center of the cell. The spatial range of the correlations is small compared to the size of the REV. Let us then consider the quantity $C_{i j}(\Delta t)$ as the spatial integral of the correlation function, i.e.,

\begin{equation}
C_{i j}(\Delta t) \equiv \int_{0}^{\infty} d \Delta x \int_{0}^{\infty} d \Delta y C_{i j}(\Delta x, \Delta y, \Delta t) \:.
\label{eq:lambda}
\end{equation}
In the non-homogeneous part of the Hele-Shaw cell (for instance close to the inlets) $C_{i j}$ would also depend on $x, y$. In the case of systems without memory, the temporal range of the correlations is also small, and it is then appropriate to also integrate over $\Delta t$:

\begin{equation}
\Lambda_{i j}^\infty \equiv \int_{0}^{\infty} d \Delta t C_{i j}(\Delta t) \:.
\label{eq:lambdaihsytar}
\end{equation}
In systems with memory the temporal range of the correlations is not small. This is discussed for instance in Ref.~\cite{Bedeaux2024a}.
It follows from microscopic reversibility and the regression hypothesis that $C_{i j}$ are symmetric matrices. 

Since the sum of saturations adds to one, there is only one independent saturation. It then follows that
\begin{equation}
\sum_{i} C_{i j}(\Delta x, \Delta y, \Delta t)=\sum_{j} C_{i j}(\Delta x, \Delta y, \Delta t)=0
\end{equation}
and for the spatially integrated form
\begin{equation}
\sum_{i} C_{i j}(\Delta t)=\sum_{j} C_{i j}(\Delta t)=0 \:.
\end{equation}
For two fluids this gives

\begin{equation}
\begin{split}
C_{w w}(\Delta x, \Delta y, \Delta t) & = -C_{w n}(\Delta x, \Delta y, \Delta t)\\
                                      & = -C_{n w}(\Delta x, \Delta y, \Delta t) = C_{n n}(\Delta x, \Delta y, \Delta t)
\end{split}
\end{equation}
and 
\begin{equation}
C_{w w}(\Delta t)=-C_{w n}(\Delta t)=-C_{n w}(\Delta t)=C_{n n}(\Delta t) \:.
\label{eq:lambdaww}
\end{equation}
These properties are true whether the fluids are compressible or not. 

\subsection{Incompressible fluids}
The mass densities for incompressible fluids in the REV are related to the saturations by

\begin{equation}
\rho_{i}^{\mathrm{REV}}(x, y, t)=\rho_{i} \phi S_{i}^{\mathrm{REV}}(x, y, t) \quad \text { for } \quad j=w, n
\end{equation}
where $\rho_{i}$ are the constant mass densities of the pure fluid phases. Furthermore we assume that the porosity $\phi$ does not depend on $x, y$. 

The mass balance is
\begin{equation}
\frac{\partial \rho_{i}^{\mathrm{REV}}(x, y, t)}{\partial t}=-\operatorname{div} \mathbf{J}_{i}^{\mathrm{REV}}(x, y, t)
\label{eq:mass_conservation}
\end{equation}
For the saturations of incompressible fluid phases this implies

\begin{equation}
\frac{\partial S_{i}^{\mathrm{REV}}(x, y, t)}{\partial t}=-\frac{1}{\rho_{i} \phi} \operatorname{div} \mathbf{J}_{i}^{\mathrm{REV}}(x, y, t)
\label{eq:incomp}
\end{equation}

For a surface area $A_{x y}=L_{x} \times L_{y}$, where $L_{x}>L_{x}^{\mathrm{REV}}$ and $L_{y}>L_{y}^{\mathrm{REV}}$, Eq.~\ref{eq:SjSrev} together with Eq.~\ref{eq:incomp} gives for incompressible fluid phases

\begin{equation}
\begin{aligned}
A_{x y}\frac{\partial S_{i}(x, y, t)}{\partial t}  = & \int_{x}^{x+L_{x}} d x^{\prime} \int_{y}^{y+L_{y}} d y^{\prime} \frac{\partial S_{i}^{\mathrm{REV}}\left(x^{\prime}, y^{\prime}, t\right)}{\partial t} = \\
 & -\frac{1}{\rho_{i} \phi} \int_{x}^{x+L_{x}} d x^{\prime} \int_{y}^{y+L_{y}} d y^{\prime} \operatorname{div} \mathbf{J}_{i}^{\mathrm{REV}}\left(x^{\prime}, y^{\prime}, t\right) = \\
 & \frac{1}{\rho_{i} \phi} \left\{\int_{y}^{y+L_{y}} d y^{\prime}\left[J_{i, x}^{\mathrm{REV}}\left(x, y^{\prime}, t\right)-J_{i, x}^{\mathrm{REV}}\left(x+L_{x}, y^{\prime}, t\right)\right]\right. +\\
& \left.\int_{x}^{x+L_{x}} d x^{\prime}\left[J_{i, y}^{\mathrm{REV}}\left(x^{\prime}, y, t\right)-J_{i, y}^{\mathrm{REV}}\left(x^{\prime}, y+L_{y}, t\right)\right]\right\}
\label{eq:dSdt}
\end{aligned}
\end{equation}
The subscripts indicate the $x$ or $y$ component of the fluxes. The aim is to substitute Eq.~\ref{eq:dSdt} into Eq.~\ref{eq:Cij} and to use the fluctuation-dissipation theorems for the mass fluxes, in order to obtain similar theorems for the fluctuation of the time derivative of the saturations. In the next section we discuss the relevant equations for the mass fluxes.

\subsection{ Fluctuation-dissipation theorems for the mass fluxes}

We restrict ourselves to the case that the temperature $T$ in the Hele-Shaw cell is everywhere constant. For this case, the linear flux-force relations for the mass fluxes with memory \cite{Bedeaux2024a} are given by

\begin{equation}
\mathbf{J}_{i}^{\mathrm{REV}}(x, y, t)=\int_{-\infty}^{\infty} \sum_{j=1}^{n} L_{ij}\left(t-t^{\prime}\right) \mathbf{X}_{j}^{q}\left(x, y, t^{\prime}\right) d t^{\prime}
\label{eq:JREV}
\end{equation}
 In this expression $\mathbf{J}_{i}^{\mathrm{REV}}$ are the mass fluxes, while $\mathbf{X}_{j}^{q}\left(x, y, t^{\prime}\right) \equiv-\frac{1}{T} \operatorname{grad} \mu_{j, T}\left(x, y, t^{\prime}\right)$ are the conjugate thermodynamic forces. The gradient of the chemical potentials is taken, keeping the temperature constant. This is indicated by the subscript $T$ on the chemical potential. Eq.~\ref{eq:JREV} gives the fluxes on the coarse-grained level. In order to describe fluctuations, we need to add a random contribution \cite{bedeaux2021,Bedeaux2024a,Bedeaux2024}
\begin{equation}
\mathbf{J}_{i, t o t}^{\mathrm{REV}}(x, y, t)=\mathbf{J}_{i}^{\mathrm{REV}}(x, y, t)+\mathbf{J}_{i, R}^{\mathrm{REV}}(x, y, t) \:,
\end{equation}
where the average of the random contribution in the last term is zero. The systematic contribution is in many cases larger, but this does not have to be so. At equilibrium it is zero, but in the experiment that we consider, it is substantial. The correlation functions are according to the fluctuation-dissipation theorems given by

\begin{equation}
\begin{aligned}
\left\langle\mathbf{J}_{i, R}^{\mathrm{REV}}(x, y, t) \mathbf{J}_{j, R}^{\mathrm{REV}}\left(x^{\prime}, y^{\prime}, t^{\prime}\right)\right\rangle =  2 \kappa_B L_{i j}\left(t^{\prime}-t\right) \mathbf{1} \delta\left(x^{\prime}-x\right) \delta\left(y^{\prime}-y\right)
\label{eq:fluctdiss}
\end{aligned}
\end{equation}
where $\mathbf{1}$ is a unit tensor. We assume the spatial correlations to be short range compared to the macroscale phenomena, which gives the $\delta\left(x^{\prime}-x\right) \delta\left(y^{\prime}-y\right)$ term. When the Hele-Shaw cell is homogeneous the matrix $L_{i j}$ is independent of $x, y$. In the actual experimental cell this is only the case away from the inlet along the central part of the model. For a stationary system $L_{i j}$ only depends on $\left(t^{\prime}-t\right)$, and not on both $t$ and $\left(t^{\prime}-t\right)$. The matrix of Onsager coefficients is symmetric. In the case without memory, this symmetry was verified using network simulations by Winkler et al.\cite{winkler2020}.

The Fourier transform of Eq.~\ref{eq:JREV} gives
\begin{equation}
\mathbf{J}_{i}^{\mathrm{REV}}(x, y, \omega)=\sum_{j=1}^{n} L_{i j}(\omega) \mathbf{X}_j^q(x, y, \omega)
\end{equation}
where the Fourier transforms are defined by 

\begin{equation}
\begin{aligned}
\mathbf{J}_{i}^{\mathrm{REV}}(x, y, \omega) & \equiv \int_{-\infty}^{\infty} e^{i \omega t} \mathbf{J}_{i}^{\mathrm{REV}}(x, y, t) d t \\
\mathbf{X}_{i}^{q}(x, y, \omega) & \equiv \int_{-\infty}^{\infty} e^{i \omega t} \mathbf{X}_{i}^{q}(x, y, t) d t \\
L_{i j}(\omega) & \equiv \int_{-\infty}^{\infty} e^{i \omega t} L_{i j}(t) d t=\int_{0}^{\infty} e^{i \omega t} L_{i j}(t) d t
\end{aligned}
\end{equation}
In the last equality we used that  $L_{i j}(t)=0$ for $t<0$ because of causality.

\subsection{Correlations for the saturation derivatives}
\label{sec:theory}
Substitution of Eq.\ref{eq:dSdt} into Eq.4 and use of Eq.\ref{eq:fluctdiss} for incompressible fluid phases gives after some algebra: 
\begin{equation}
\begin{aligned}
C_{i j}(\Delta x, \Delta y, \Delta t) \equiv \left\langle\frac{\partial S_{i}(x, y, t)}{\partial t} \frac{\partial S_{j}(x+\Delta x, y+\Delta y, t+\Delta t)}{\partial t}\right\rangle  = \frac{8 \kappa_B L_{i j}(\Delta t)}{\rho_{i} \rho_{j} \phi^{2} A_{xy}} \delta(\Delta x) \delta(\Delta y)
\label{eq:Cij2}
\end{aligned}
\end{equation}

\begin{figure*}
	\centering
	\includegraphics[width=\linewidth]{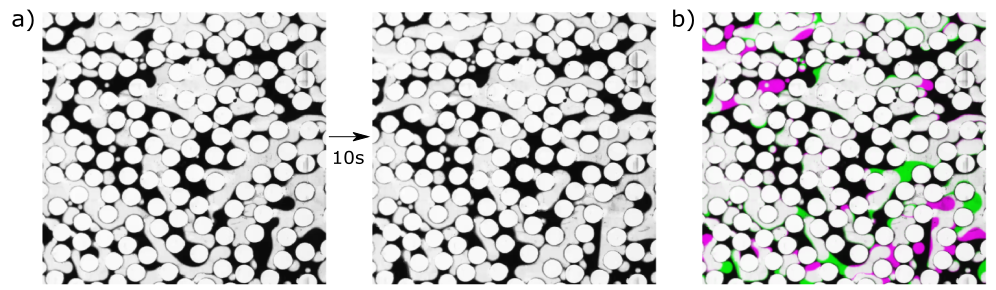}
	\caption{Fluctuation in phase saturation in the porous medium. The diameter of the cylinders is $2$mm. In a) we see two snapshots of the experiment separated by 10s. On average, the flow is from left to right. In b) we see a composition of the two snapshots highlighting the regions where fluid reconfiguration has occurred. In green we see points that were oil-filled and became water-filled and in magenta points that were water-filled that became oil-filled.}
	\label{fig:AV_fluctuation}
\end{figure*}
Equation \ref{eq:lambda} subsequently gives
%
\begin{equation}
C_{i j}(\Delta t)=  \int_{0}^{\infty} d \Delta x \int_{0}^{\infty} d \Delta y C_{i j}(\Delta x, \Delta y, \Delta t)  
=  \frac{8 \kappa_B L_{i j}(\Delta t)}{\rho_{i} \rho_{j} \phi^{2} A_{xy}}
\label{eq:lambdaint}
\end{equation}

We shall use this to obtain the conductivity matrix $L_{i j}$ from the fluctuations of the time derivative of the saturation. The corresponding Fourier transform is
\begin{equation}
C_{i j}(\omega)=\frac{8 \kappa_B L_{i j}(\omega )}{\rho_{i} \rho_{j} \phi^{2} A_{xy}} \:.
\label{eq:lambdaomega}
\end{equation}
This equation applies to all frequencies, but in this work we will consider only the zero frequency component. With Eq.~\ref{eq:lambdaww} it follows that
\begin{equation}
\frac{L_{w w}}{\rho_{w}^{2}}=-\frac{L_{w n}}{\rho_{w} \rho_{n}}=-\frac{L_{n w}}{\rho_{n} \rho_{w}}=\frac{L_{n n}}{\rho_{n}^{2}}
\label{eq:Lww}
\end{equation}
This equation means that if we measure one coefficient, we know all of them. 
This property of the Onsager matrix follow from the relationship imposed upon the fluxes. There is only one flux-force product contributing to the entropy production (energy dissipation), because there is only one driving force. Any extended expression in terms of two fluxes (still only one driving force), leads to the inter-dependency of coefficients. A gradient in saturation gives a second driving force. Without this, an expansion of the flux description to describe two-component flow, will give an undetermined set of coefficients. Equation \ref{eq:Lww} is valid as a function of $\Delta t$ and as a function of $\omega$. But the matrix of the $L_{i j}$-coefficients can not be inverted, as the determinant of the coefficient matrix is zero.  Nevertheless, the various coefficients are included here because they can be used to make a link to the common permeability model, an idea we are currently working on, a preliminary version of which can be found in the Appendix. 

As previously mentioned, in the case of systems without memory in which the temporal range of correlations is small, it is convenient to integrate over $\Delta t$. From Eqs.~\ref{eq:lambdaint} and \ref{eq:lambdaihsytar} we have:
\begin{equation}
\Lambda_{i j}^\infty = \int_0^\infty d\Delta t C_{i j} (\Delta t) = \frac{8 \kappa_B}{\rho_{i} \rho_{j} \phi^{2} A_{xy}} \int_0^\infty d\Delta t L_{i j}(\Delta t)  \:.
\label{eq:lambdaintnomem0}
\end{equation}
In this limit of fast decaying temporal correlations, we can make the assumption $L_{i j}(\Delta t) = L_{i j}\delta (\Delta t)$. We then have
\begin{equation}
\Lambda_{i j}^\infty = \frac{8 \kappa_B L_{i j}}{\rho_{i} \rho_{j} \phi^{2} A_{xy}}   \:.
\label{eq:lambdaintnomem}
\end{equation}

Using this equation, the product  $\kappa_B L_{i j}$ can be obtained from the measurement of $\Lambda_{i j}^\infty$. We will later see that the measurement of $\Lambda_{i j}^\infty$ itself presents some challenges and we will show how a related quantity $\Lambda_{i j}^*$ can be used instead. We have focused on the wetting phase saturation $S_w$, from which we aim to access $L_{ww}$ as discussed. The other coefficients can be obtained in the case of incompressible fluids via Eq.~\ref{eq:Lww}. A different approach was taken recently by Alfazazi et al.~\cite{Alfazazi2024} where the authors considered the full volume flux composed of both fluid phases instead, as a pathway to recover the Onsager coefficient. 

\section{Results and Discussions}
\label{sec:results}
\subsection{Experimental dynamics, phase redistribution and fluctuations in saturation}

Using the setup shown in Fig.~\ref{fig:syst}, we performed an experiment in which both water and mineral oil phases were co-injected at the fixed injection rate $q$. If the injection rate was too low the experiment would take a very long time to reach a steady-state situation and the fluid transport would be dominated either by connected pathways linking the inlet to outlet or by flow through thin films covering the surface of the grains \cite{aursjo2014}. These dynamic states would look frozen to the camera, i.e., we would not see a clear dynamics of moving clusters and there would be no observable temporal fluctuations of saturation, something needed for the computation proposed in Eq.~\ref{eq:Cij2}. On the other hand, if the injection rate were set to a very high value, we would require a very high acquisition rate from the cameras to capture the temporal dynamics at a rate fast enough to perform the saturation analysis. By trial and error, we found the rate of $q = 16$ ml/h to be well suited for our purposes. The capillary number can be estimated as $Ca = \eta_o q/(A^x\gamma)$ where $\gamma$ is the oil-water interfacial tension and $A^x = 35mm^2$ is the cross sectional area of the model. We have not measured $\gamma$ for the particular fluid pair in the system, but using a typical value of $\gamma ~\approx 20mN/m$ would give  $Ca \approx 1.4\: 10^{-4}$. The flow is thus in a domain where viscous forces are expected to be rather significant.

\begin{figure}[t]
	\centering
	\includegraphics[width=0.5\linewidth]{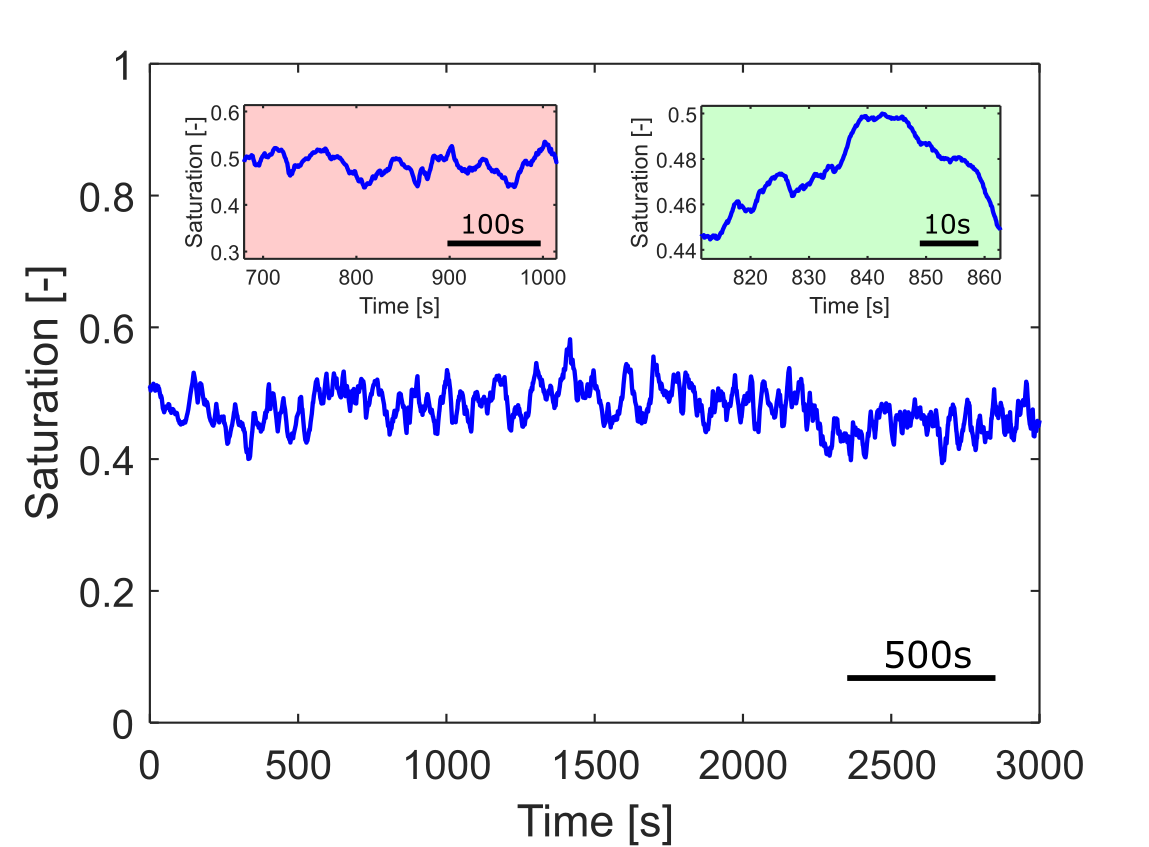}
	\caption{Water phase saturation as a function of time. In the main plot we show the full time series of the water phase saturation in the region analyzed. We see that as expected this quantity fluctuates around a mean value which here is measured to be around $\overline{S_w}=0.48$. The scale bar for this plot shows 500 seconds and we can see that at this time scale, the fluctuations seem uncorrelated. On the inset at the top left (light red), we see a zoomed in section showing what the typical fluctuations look like at the time scale of 100 seconds. Here again we see that the fluctuations seem uncorrelated. On the top right (light green) we see a further zoomed in section showing fluctuations ocurring at the tipical time scale of 10 seconds. At this magnification level, we expect the fluctuations to be highly correlated.}
	\label{fig:saturation_plot}
\end{figure}

The experiment is initiated with the model completely saturated by the water phase. The two water and oil pumps are then simultaneously turned on and the dynamics begins. During the experiment, an intense secession of events occur, with fluid clusters getting trapped and mobilized, merging and breaking apart. This dynamics causes the local saturation in any portion of the medium to fluctuate in time, and this fluctuation is the basic quantity we measure. In Fig.~\ref{fig:AV_fluctuation} we see the typical fluid reconfigurations that lead to fluctuations in saturation (images from the machine vision camera focused on the red square part seen in Fig.~\ref{fig:syst}). Figure~\ref{fig:AV_fluctuation}a shows two snapshots of the experiment separated by 10s (100 frames). The average flow direction is from left to right. Fig.~\ref{fig:AV_fluctuation}b) shows a composition of the two snapshots where we highlight the portions where fluid reconfiguration has occurred. In green we see points that were oil-filled and became water-filled and in magenta points that were water-filled that became oil-filled. Notice that since the boundaries of the image are all open to the flow, nothing prevents us from having more green or more magenta in the composition image, i.e., the total saturation of each fluid phase can vary. If the saturation of phase $i$ in the region we consider increases by a given amount $\delta S_i$, the saturation of the other phase must decrease by the same amount (as $S_{w}^{\mathrm{REV}}+S_{n}^{\mathrm{REV}}=1$), however, it is important to remember that the individual values of $\delta S_i$ do not need to be zero. In a steady-state situation, we expect the time average of the fluctuations in phase saturation to vanish, i.e., $\left<\delta S_i\right> = 0$, and the saturation fluctuates around a well defined mean value $<{S_i}>$, however the specific value of $\delta S_i$ at a given time does not need to vanish.

Figure~\ref{fig:saturation_plot} shows a plot of the water saturation in the REV as a function of time for an interval of 3000 seconds (50 minutes). In the very beginning of the experiment (not seen in the plot), the model was fully saturated with water (thus having water saturation equal to 1). In the analysis shown here, we waited for a period of about 24 minutes since the beginning, for the saturation to decrease to the steady-state value we see in the plot, in which it fluctuates around the mean value $<{S_i}> = 0.48$. We start counting the time at this point, when the system is already in a steady-state. In addition to the main plot (with a scale bar showing 500 seconds), we show two insets. On the top left (light red), we show the typical fluctuations at a scale of 100 seconds and on the top right (light green), we show the fluctuations at a scale of 10 seconds. Just from the visual inspection of the plot, we can expect the fluctuations to be uncorrelated on time scales larger than 100 seconds, but within the 10 seconds time scale, the fluctuations seem to be strongly correlated. What we mean by that is that if we know that at a given instant the saturation is, say, increasing, we can say that there is a high likelihood that it will also be increasing within the next few seconds, but we cannot say anything about it 100 seconds into the future.

\begin{figure}
	\centering
	\includegraphics[width=0.5\linewidth]{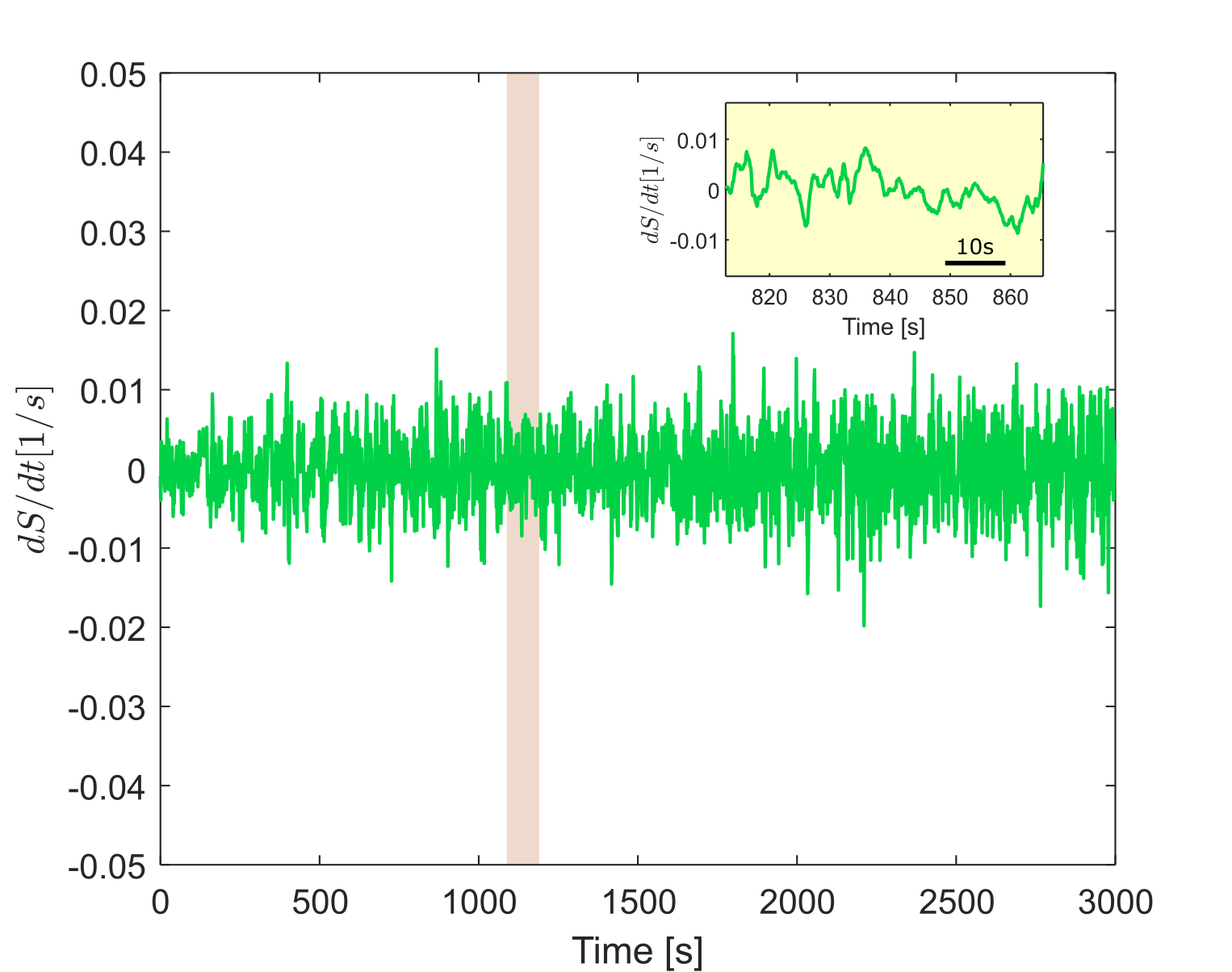}
	\caption{Derivative of the water saturation as function of time. As expected, the fluctuations in saturation have zero average, this is a necessary condition for a steady-state system. On the inset (light yellow), we show a magnified portion of about 50 seconds in the plot which indicates that indeed the signal seems to be uncorrelated for times larger than 10 seconds. The vertical band on the main plot denotes a sliding window of extent $\Delta t^{sub} = 100$ s which will become relevant for the analysis in Section~\ref{sec:ensemble}.}
	\label{fig:saturation_derivative_plot}
\end{figure}

Our next step in the study of the fluctuation correlations is to consider the derivative of the saturation. Taking a numerical derivative of a strongly fluctuating experimental quantity (as seen in Fig.~\ref{fig:saturation_plot}) can be challenging, particularly for very short time scales in which high-frequency fluctuations (possibly coming from other sources, like fluctuations in illumination or electronic noise) can lead to overestimating the result. In order to account for that, we have considered a time difference $\delta t$ for the finite difference used to compute the derivative that is in between the minimum time scale we can resolve of $0.1$ seconds (corresponding to the time difference between two frames in the images) and the typical time scale we expect the data to start presenting decorrelation in the signal (which we conservatively estimate as $10$ seconds from Fig.~\ref{fig:saturation_plot}). Attending to the condition $0.1s \ll \delta t \ll 10s$ we chose $\delta t = 1s$. Fig.~\ref{fig:saturation_derivative_plot} shows the resulting plot of the derivative of the water saturation. As expected, this quantity fluctuates nicely around zero. This is expected from the fact that the experiment is in a statistical steady-state as previously explained. On the inset (light yellow) we see a zoomed in section indicating the expected fluctuations in the derivative of saturation for a typical time scale of 10 seconds. We see now more clearly that indeed the decorrelation time seems to be smaller than 10 seconds. Under these conditions, the system is in a state which can be characterized by microscopic reversibility.

\begin{figure}[t]
	\centering
	\includegraphics[width=0.5\linewidth]{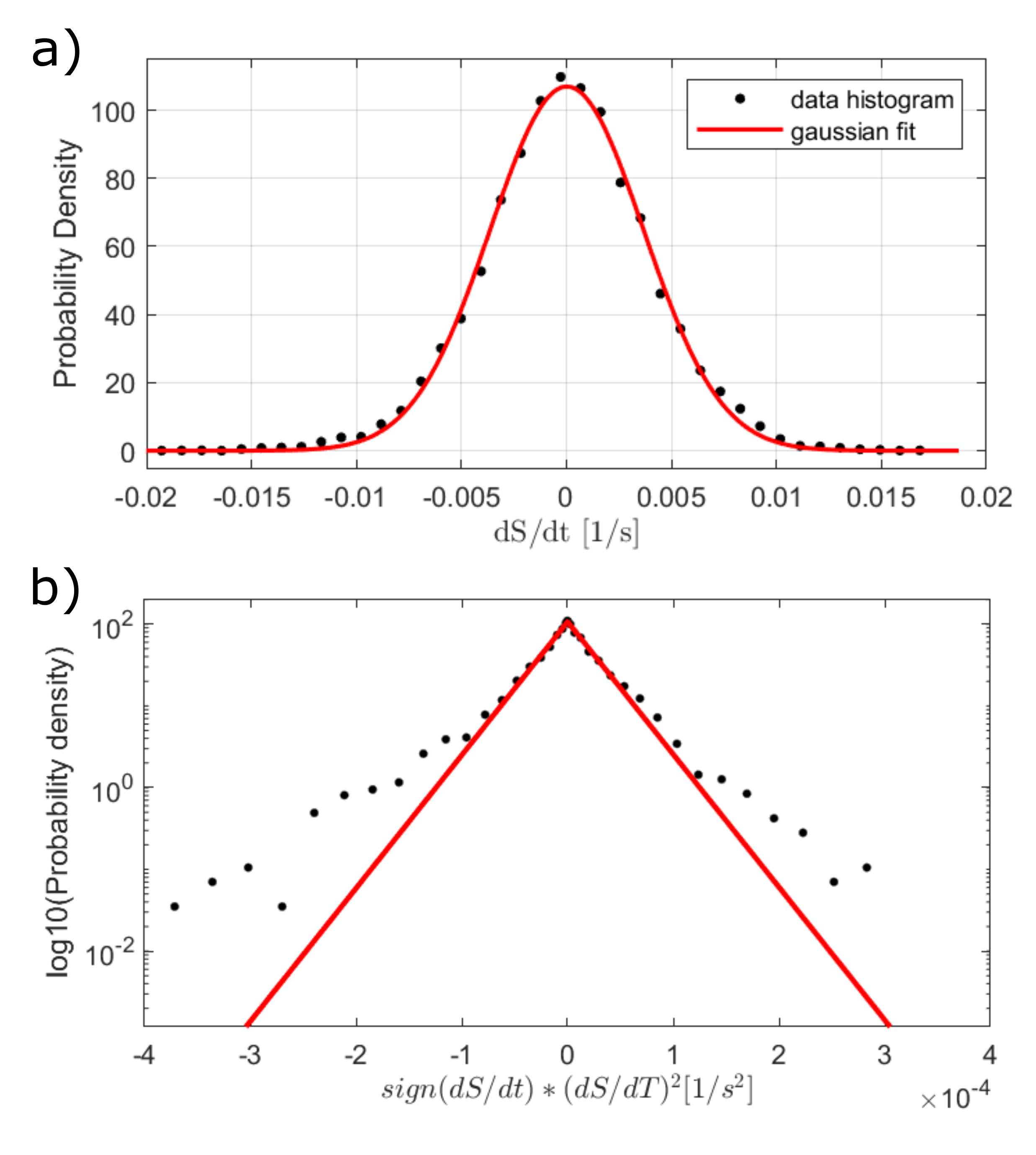}
	\caption{Statistics of the fluctuations in the derivative of the water saturation (data in Fig.~\ref{fig:saturation_derivative_plot}). The dots show the probability density function calculated from a histogram of the data. The solid line is a Gaussian fit to the data. In a) we show the data on a linear scale and in b), we plot the probability on a semilog scale as function of $sign(dS/dt)*(dS/dT)^2$. Although the Gaussian fit seems very well suited from the linear plot in a), we see in b) that the few rare events on the tails of the distribution seem to deviate from the Gaussian prediction.}
	\label{fig:gaussian}
\end{figure}

We further quantify the fluctuations by looking at their statistics. In Fig.~\ref{fig:gaussian} we show the probability density function of the fluctuations $P(x=dS/dt)$ obtained by computing a histogram of the data in Fig.~\ref{fig:saturation_derivative_plot}. The solid line shows a Gaussian fit of the form $P(x) = A \exp (-(x-x_0)/2\sigma^2)$ with parameters $A = 106.9 $ s, $x_0 = 1.51 \:10^{-5}$ [1/s] and $\sigma =  0.0508 $ [1/s]. Notice that the curve is essentially centered at $dS/dt = 0$, another indication that the system is in a steady-state and the fluctuations have zero temporal average (we did not subtract the mean value of the fluctuations prior to producing the histogram in Fig.~\ref{fig:gaussian}, the zero mean is a natural consequence of the steady-state dynamics). In Fig.~\ref{fig:gaussian}a we show the data on a linear scale and the Gaussian fit seems to describe the data reasonably well. However, when plotting the probability as function of the reduced variable sign$(dS/dt)\cdot(dS/dT)^2$, see Fig.~\ref{fig:gaussian}b, we notice that the rare events of large saturation changes (with $|dS/dt|>0.01$) do not seem to obey the Gaussian statistics. Although these events are relatively unfrequent (they are about 2 standard deviations away from the mean, therefore accounting for less than $5\%$ of the total count of events), we believe they could be very significant for the process as those should correspond to the large cluster rearrangements leading to fast changes in saturation. We found however that this result (the apparent deviation from Gaussian statistics) was particularly sensitive to the interval $\delta t$ used in the computation of the derivatives. Whether or not this deviation is significant is still an open question and more research is needed to further clarify this point.

\subsection{Autocorrelation of the saturation derivative}
\label{sec:autocorr}

Having analyzed the statistics of the fluctuations of the derivative of the saturation, we move now to the analysis of the autocorrelation $C_{ww}$ of the signal. The autocorrelation function is given by

\begin{equation}
\begin{aligned}
C_{ww}(\Delta t) \equiv\left\langle\frac{\partial S_{w}(t)}{\partial t} \frac{\partial S_{w}(t+\Delta t)}{\partial t}\right\rangle = & \frac{1}{\tau} \int_{t}^{t+\tau} d t^{\prime} \frac{\partial S_{w}\left( t^{\prime}\right)}{\partial t^{\prime}} \frac{\partial S_{w}\left(t^{\prime}+\Delta t\right)}{\partial t^{\prime}} = \\ & \frac{1}{\tau} \int_{0}^{\tau} d t^{\prime} \frac{\partial S_{w}\left(t^{\prime}\right)}{\partial t^{\prime}} \frac{\partial S_{w}\left(t^{\prime}+\Delta t\right)}{\partial t^{\prime}} \:.
\label{eq:Cww}
\end{aligned}
\end{equation}
Notice that in the last line of Eq.~\ref{eq:Cww} we made the assumption that the autocorrelation function will depend only on the time lag $\Delta t$ and not on the time $t$ itself. This assumption is reasonable here as we have shown that the experiment is on a statistical steady-state and any given time $t$ is statistically similar to any other. The steady-state assumption and this idea that, in a statistical sense, any time is just as good a starting point as any other will prove useful later in Section \ref{sec:ensemble}.

In the theory section we have considered correlations between space and time, i.e., $C_{ww}(\Delta x, \Delta y, \Delta t)$ but in the experiments we will focus only on the temporal correlations. Our function $C_{ww}(\Delta t)$ can then be understood as the spatially integrated correlation from Eq.~\ref{eq:lambda}. We believe this assumption to be reasonable given that we are restricting out attention to the region of the experiment in a statistical steady-state where the flows are spatially homogeneous on a scale larger than the REV size. It has been shown that close to the inlets of the system, the flows are not spatially homogeneous and strong channelization can occur \cite{aursjo2014}. The study of spatial correlations in such systems is important as it can give information about how far from the boundaries the steady-state is achieved, but this topic falls outside the scope of the present work.

\begin{figure}[htbp]
	\centering
	\includegraphics[width=0.5\linewidth]{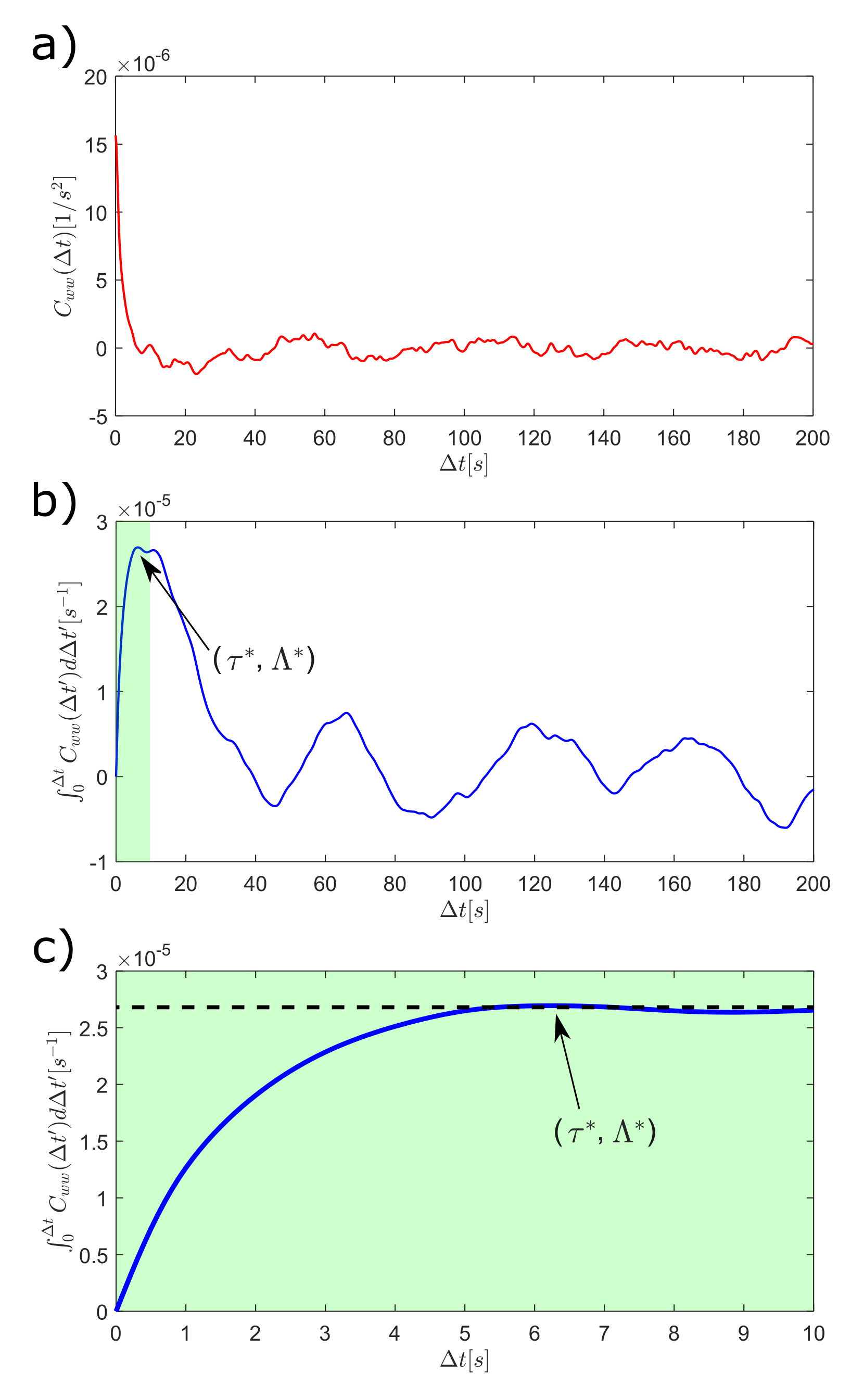}
	\caption{a) Autocorrelation function $C_{ww}(\Delta t)$ of the derivative of the water saturation (data in Fig.~\ref{fig:saturation_derivative_plot}). Notice that, as the data loses its correlation in less than 10 seconds, $C_{ww}(\Delta t)$ fluctuates around zero for large values of the time lag $\Delta t $. b) Integral of the autocorrelation function $C_{ww}(\Delta t)$. If $C_{ww}(\Delta t)$ decayed exponentially, we would expect its integral to grow asymptotically. The arrow denotes the first peak in the integral, having coordinates $\left(\tau^* = 6.2 s, \Lambda^* = 2.6938 \: 10^{-5} s^{-1}\right)$ and we will show later that this is a good proxy for the expected asymptotic behavior that would be observed without the spurious effects from the uncorrelated region of the graph for large time lag $\Delta t $. c) shows a zoomed in section for the region highlighted in green in b), i.e., for times not much larger than $\tau^*$. The dashed line indicates the schematically the plateau level that one would expect without the disturbance from the uncorrelated regions for large $\Delta t $.}
	\label{fig:autocorr}
\end{figure}

\begin{figure}[t]
\centering
	\includegraphics[width=0.5\linewidth]{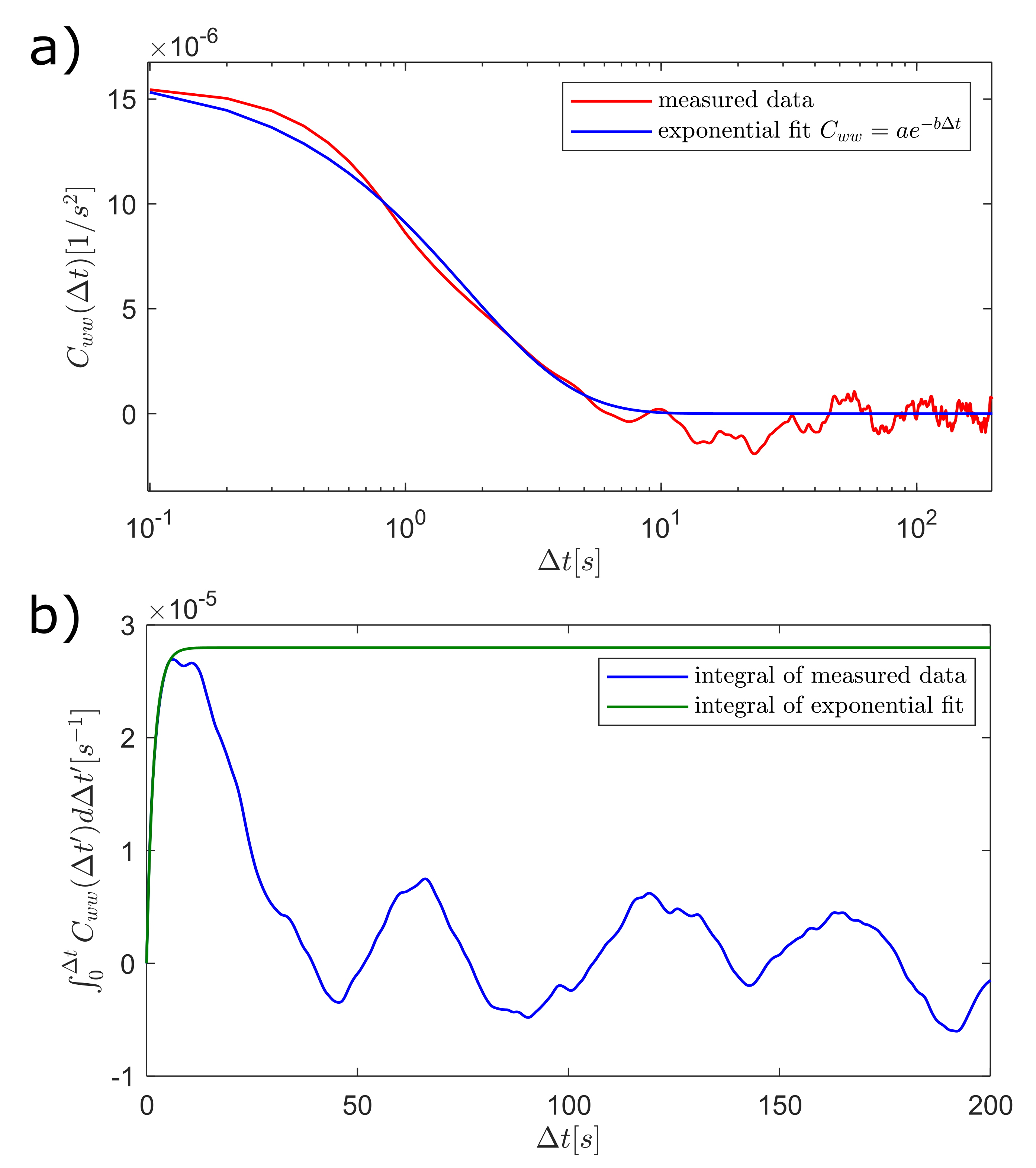}
	\caption{a) Exponential model (blue) fitted to the experimental data (red) of the autocorrelation function presented on a semilog scale. The fitting is produced using the data restricted to the interval $\Delta t < \tau^*$ b) Integral of the autocorrelatuon function of the measured data (blue) and the exponential model (green). The plateau value from the fitted model is measured to be $\Lambda^*_{fit} = -a/b = 2.799 \: 10^{-5} s^{-1}$. This value is only $4\%$ higher than the previously measured value of $\Lambda^* = 2.6938 \: 10^{-5} s^{-1}$ calculated using the position of the first peak in the integral.}
 \label{fig:autocorr_ensemble2}
\end{figure}

The autocorrelation function is shown in Fig.~\ref{fig:autocorr}a). Here we see that indeed, as expected, for large values of the time lag $\Delta t $, the signal gets completely uncorrelated. Experimentally however, we do not measure a clear exponential decay but observe a somewhat oscillating signal around the zero level for large values of $\Delta t $. Similar observations were also previously reported \cite{Alfazazi2024}. Consider now the integral of the autocorrelation function

\begin{equation}
   \Lambda(\Delta t) = \int_{0}^{\Delta t} C_{ww}(\Delta t ^\prime )d\Delta t ^\prime \:.
    \label{eq:Lambda_exp}
\end{equation}
If the autocorrelation function decayed exponentially to zero, we would expect its integral to grow towards an asymptotic plateau $\Lambda^{\infty} = \Lambda(\Delta t = \infty)$, see Eq.~\ref{eq:lambdaintnomem}. However, in the case experimentally measured here, due to our limited statistics, the integral does not seem to approach a plateau for large  values of $\Delta t$ as can be seen in Fig.~\ref{fig:autocorr}b). The uncorrelated region of the data generates a sequence of valleys and hills that extends far into the region of large values of $\Delta t$. We notice however that this integral presents a well defined first peak, marked by the black arrow in Fig.~\ref{fig:autocorr}b). The position of this peak here has coordinates $\left(\tau^* = 6.2 s, \Lambda^* = 2.6938 \: 10^{-5}s^{-1}\right)$. In c) we show a zoom in on the area highlighted in green in b), i.e., for very sort time lags in the region where the signal has not yet completely decorrelated. If we focus our attention on this area of the signal, we can indeed see an indication of the expected plateau level $\Lambda^{\infty}$ shown schematically by the dashed line in the figure. Another potential way of performing the measurement would be to first fit an exponentially decaying curve of the form $C_{ww}=a e^{-b\Delta t}$ to the autocorrelation function and then integrate from the fit. The result is shown in Fig.~\ref{fig:autocorr_ensemble2} where we show in a) the fitting in a semilog plot and in b) its integral. The plateau value obtained from the fitted curve is $\Lambda^*_{fit} = -a/b = 2.799 \: 10^{-5} s^{-1}$, which is only $4\%$ larger than our previous estimate for $\Lambda^* = 2.6938 \: 10^{-5}s^{-1}$ from the first peak position. Although this procedure seems more intuitive, as the simple exponential is a reasonable choice for the fit given that we are interested only in the zero frequency contribution to the autocorrelation for systems without memory, we found this method to be too sensitive to the domain of the data used for the fitting. In the data shown in Fig.~\ref{fig:autocorr_ensemble2} we used the time up to the estimated correlation time $\tau^* = 6.2 s$ for the fitting. However, if we had used a larger time, for instance up to $10\tau^* = 62s$, the fit would be considerably worse and the estimated plateau value would be $\Lambda^*_{fit} = -a/b = 3.012 \: 10^{-5} s^{-1}$, which is now $12\%$ larger than our previous estimate for $\Lambda^*$. Taking that into account, we decided to avoid any type of curve fitting to the data here and focus instead on using the value of the first peak of the raw signal $\Lambda^*$ as a proxy for the plateau level. In Sec.~\ref{sec:ensemble} we will show further evidence of how $\Lambda^*$ can be used as a reasonable estimate for the plateau value $\Lambda^\infty$ while still avoiding the need for fitting the data.

\subsection{Limited statistics and measurement stability}

\begin{figure}[t]
\centering
	\includegraphics[width=0.5\linewidth]{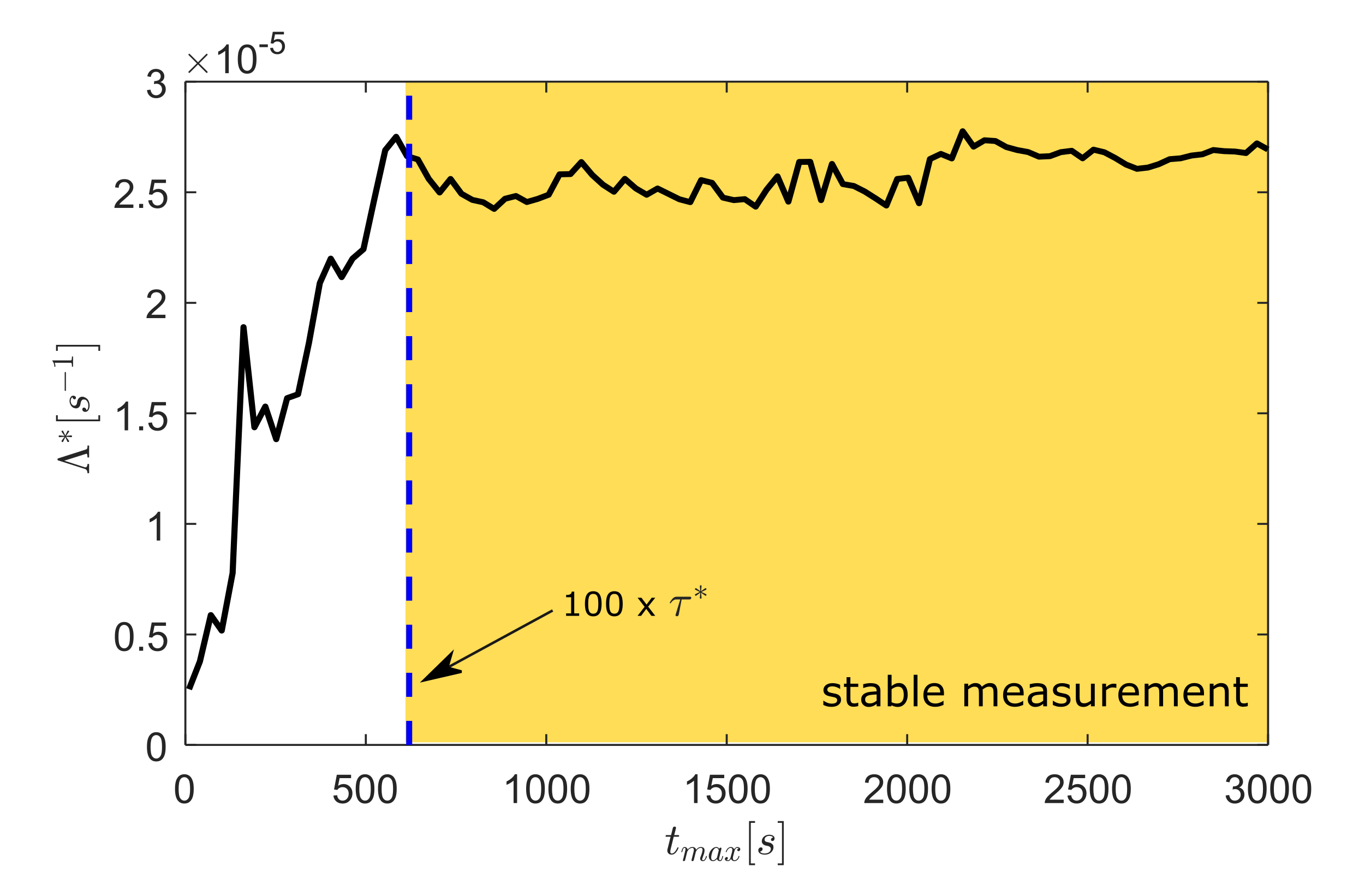}
	\caption{Dependence of $\Lambda^*$ on the duration of the experiment $t_{max}$. We have noticed that if the dataset is not long enough, the value of $\Lambda^*$ is not stable. Only for datasets longer than at least 100 times the decorrelation time $\tau^*$ we could obtain a  stable measurement for $\Lambda^*$. The yellow window shows the interval of stability.}
	\label{fig:changing_duration}
\end{figure}

Another key point that needs to be considered is the total duration of an experiment necessary to yield a statistically relevant measure for $\Lambda^*$. One can always produce the integral of the autocorrelation function and obtain a value for $\Lambda^*$, but if the experiment does not last long enough, this value is not statistically significant and a different result can be obtained by simply making the experiment last longer. In order to test that, we considered again the original dataset shown in Fig.~\ref{fig:saturation_derivative_plot} and chopped it into smaller intervals of length $t_{max}$ ranging from $t_{max}=10s$ to the duration of the full dataset with $t_{max}=3000s$. For each such subinterval we repeated the procedure of computing $\Lambda^*$ by computing the first peak in the integral of the autocorrelation function as done in Fig.~\ref{fig:autocorr}. The result is shown in Fig.~\ref{fig:changing_duration}. We see that if the experiment does not last longer than a minimum threshold (here estimated to be on the order of 100 times the decorrelation time $\tau^*=6.2$), the measurement of $\Lambda^*$ is not stable. The window marked in yellow shows the zone of stability where $\Lambda^*$ does not vary much by increasing the duration of the experiment.

In the next section we will investigate the matter of the limited statistics in our signal in more detail. We will show that $\Lambda^*$ gives indeed a reasonable proxy for the measurement of the expected plateau level $\Lambda^{\infty}$ by following a procedure inspired in a somewhat similar measurement performed numerically in \cite{winkler2020}. In that work, the authors overcame the issue of the poor statistics in the computation of the integral of the autocorrelation function by repeating the measurement for an ensemble of numerical experiments and averaging over the results. Since experimentally we do not have the capacity to repeat the experiment a very large number of times, we will rely on a different approach to study how the measurements are affected by the limited statistics and show how the plateau in the integral of the autocorrelation function is revealed as the quality of the statistics improves.

\subsection{Alternative procedure: averaging over subexperiments}
\label{sec:ensemble}

From the measurement in Fig.~\ref{fig:autocorr} we saw that the system studied seemed to lose its correlation for a time much shorter than the duration of the analysis. Indeed, if we take the time scale $\tau^* = 6.2 s$ of the first peak in the integral of the autocorrelation function (see Fig.~\ref{fig:autocorr}b) as an estimate of the typical decorrelation time, we see that this time is much shorter than the total duration $\Delta t^{tot}$ of the analysis (in the present case $\Delta t^{tot} = 3000 s$). Taking this fact into consideration, we propose to subdivide the original experimental signal shown in Fig.~\ref{fig:saturation_derivative_plot} into a large set of shorter subexperiments, each having a fixed duration $\Delta t^{sub}$ following the conditions $\Delta t^{sub} \ll \Delta t^{tot}$ and $\Delta t^{sub} \gg \tau^*$. Here we have chosen $\Delta t^{sub} = 100 s$ as a compromise between the two time scales. We can then choose any point in the original time signal from Fig.~\ref{fig:saturation_derivative_plot} and take that time step as a starting point for a new subexperiment with duration $\Delta t^{sub}$. We then perform the same analysis as before in this subexperiment and compute the autocorrelation function of the signal and its integral. Finally, we repeat this procedure for a large number of such subexperiments and average the results. This effectively results to artificially creating a pseudo-ensemble of subexperiments, over which we will average the results similarly to what was done numerically in \cite{winkler2020}. This creates a REV with properties suitable for the present analysis. We have selected 100 equally spaced time steps in the original data set and performed the analysis on each of them, thus yielding 100 subexperiments each having duration $\Delta t^{sub} = 100 s$ (see the vertical sliding window in Fig.~\ref{fig:saturation_derivative_plot} exemplifying the possible positioning of one of those subexperiments in the full dataset). Notice that since the total length of the original data set is $\Delta t^{tot} = 3000 s$, there is overlap between the time spans of the 100 sub-experiments.

\begin{figure}[H]
	\centering
	\includegraphics[width=0.5\linewidth]{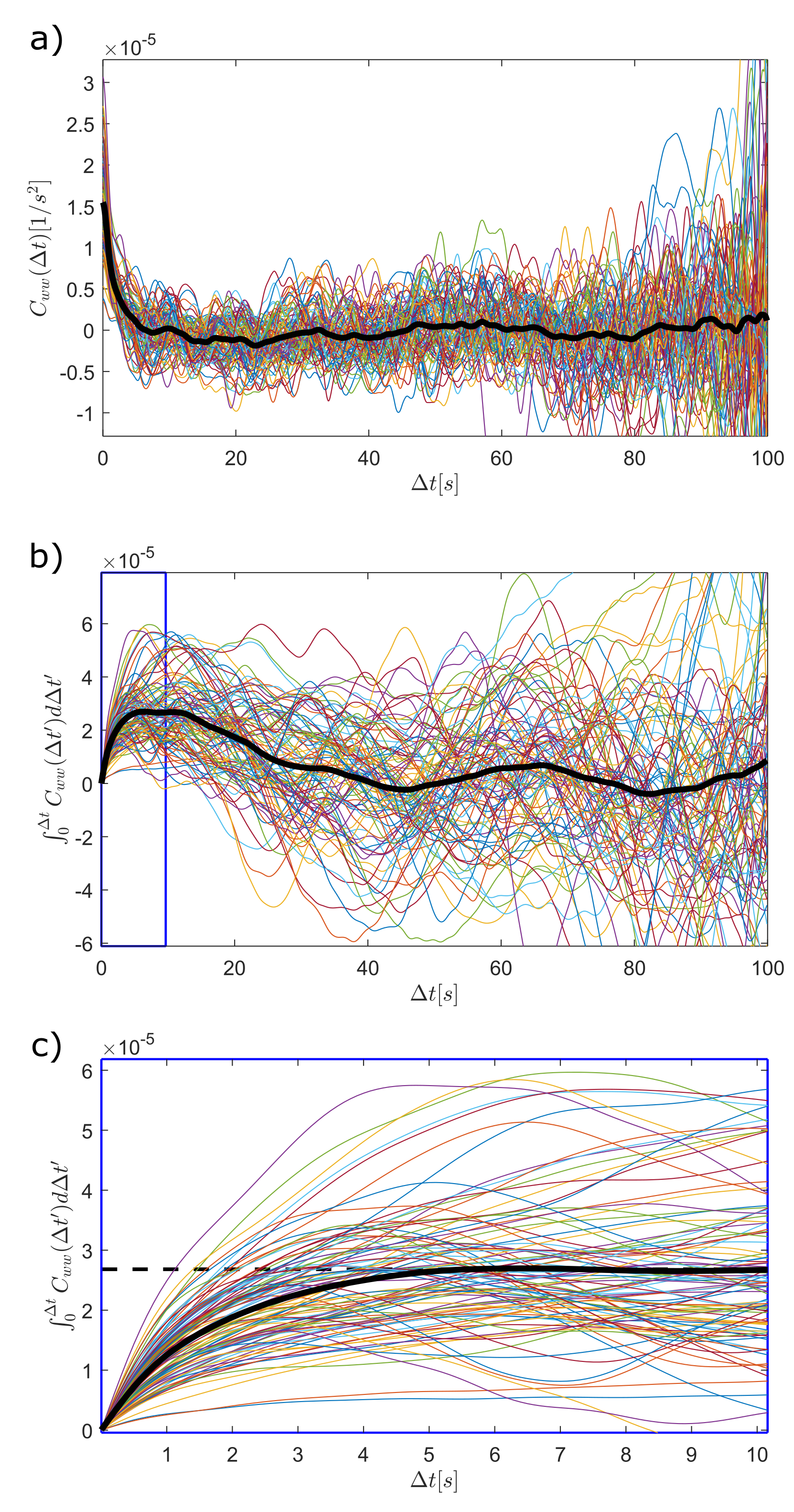}
	\caption{a) Autocorrelation function $C_{ww}(\Delta t)$ of the derivative of the water saturation (data in Fig.~\ref{fig:saturation_derivative_plot}) computed for 100 subexperiments with duration $\Delta t^{sub} = 100 s$. Each of the colored curves in the background corresponds to a subexperiment. The thick black curve is the average of all the subexperiments. Notice that, when seen individually, it would not be possible to observe the clearly decaying trend of the autocorrelation in each of the subexperiments data. It is only when we take the average of all of them that we see the decaying correlation. On b) we show the integral of the autocorrelation function. Again, the data from each subexperiment seems to vary dramatically, but when we average out all of them, we can see the formation of the plateau level for short times. This is shown more explicitly in c) where we show a zoomed in section in the blue box in b). The peak of the black curve occurs at $\left(\tau^* = 6.5 s, \Lambda^* = 2.6965 \: 10^{-5} s^{-1}\right)$ which is nearly indistinguishable from the measurement shown in Fig.~\ref{fig:autocorr} where we had obtained $\left(\tau^* = 6.2 s, \Lambda^* = 2.6938 \: 10^{-5} s^{-1}\right)$ using the full data set instead of the averaging over subexperiments presented here. Once again the dashed line indicates schematically the plateau level of $\Lambda^\infty$.}
	\label{fig:autocorr_ensemble}
\end{figure}

In Fig.~\ref{fig:autocorr_ensemble} we see the results of such measurements. The individual colored curves in the background show the measurements performed for each subexperiment and the thick black line corresponds to the pseudo-ensemble average. In the individual subexperiments, we see vastly oscillating curves that, taking individually, do not seem to present any clear trend. However, when taking the average of the data, we see the decaying of the autocorrelation function in a) and the development of the expected plateau in the integral shown in b), a result very similar to the one obtained previously in the analysis of the autocorrelation of the full data set in Fig.~\ref{fig:autocorr_ensemble}. The position of the peak of the average curve in Fig.~\ref{fig:autocorr_ensemble}b gives the estimate for the decay time $\tau^*$ and plateau level $\Lambda^*$ as $\tau^* = 6.5 s$ and $\Lambda^* = 2.6965 \: 10^{-5} s^{-1})$, again very similar to the previous values of $\tau^* = 6.2 s$ and $\Lambda^* = 2.6938 \: 10^{-5} s^{-1}$. The fact that we obtained the same result by the two separate procedures (either considering the average of several subexperiments or by a single much longer full experiment) is interesting, but not entirely surprising: the autocorrelation for each subexperiment involves an average of time lags inside the time-frame of that subexperiment but when considering the averaging over different subexperiments, we retrieve the same information as that encoded in Eq.~\ref{eq:Cww} when time averaging involves the full data set.

In spite of that, notice that, although the alternative analysis method presented here (taking the average of subexperiments) does not in principle provide any advantage as to the accuracy of the measurement (the black curves in Fig.~\ref{fig:autocorr_ensemble} are very similar to the results from Fig.~\ref{fig:autocorr}) it provides us with an insight into how this type of measurement requires a long dataset to produce statistically meaningful results, thus helping to define the conditions for a usable REV. If the data collection had not occurred for a long enough interval, say if we had only the data corresponding to a $100 s$ acquisition, i.e., the duration of one of the subexperiments seen in Fig.~\ref{fig:autocorr_ensemble}, we could measure any of the curves in the background and would not identify the plateau level in Fig.~\ref{fig:autocorr_ensemble} c). This means that for the measurement of the plateau level $\Lambda^*$ to be reliable, one really needs to have a data set that spans a very long interval. Roughly speaking, we can say that a minimum time interval of about 100 times the decorrelation time $\tau^*$ is necessary for a reliable measurement of the plateau level $\Lambda^*$, as shown in Fig.~\ref{fig:changing_duration}.

Notice that the idea of studying the individual subexperiments separately and considering their subsequent average resonates with the idea that the (ergodic) system is in a statistical steady-state and that (again, in a statistical sense) each time step is just as good as any other as a starting point for the analysis. Had the system been in a transient state (for instance, for the discarded time frames right at the beginning of the experiment, when the system is flushed with water and the co-injection of both phases starts), this idea would not be applicable.

\subsection{Changing the area of the observation window}

In the previous sections, we have seen two equivalent methods of computing the plateau level of the integral of the autocorrelation function, Eq.~\ref{eq:Lambda_exp}. One of the methods could be interpreted as a temporal average of a long experiment, and the other as a pseudo-ensemble average of several shorter subexperiments (each of them involving also a temporal averaging for the computation of each colored curve seen in the background of Fig.~\ref{fig:autocorr_ensemble}). The two methods yielded the same value for estimate $\Lambda^*$ of the asymptotic value of the integral. In particular, although the averaging was different, the area of the observation window in the $xy$ plane of the experiment, $A_{xy}$, was the same for both, i.e., we used the full frame of view of the camera to collect the images (See Fig.~\ref{fig:AV_fluctuation}). From the theoretical analysis presented in Sec.~\ref{sec:theory}, we see that the value of $\Lambda^*$ is given by Eq.\ref{eq:lambdaomega}, i.e.,  $\Lambda^*=8 \kappa_B L_{w w}/\left(\rho_{w}^{2} \phi^{2} A_{xy}\right)$. Notice in particular that this equation predicts that if the parameter $\kappa_B$ (theorized to be an yet unspecified numerical constant) is indeed invariant over the experiments, there is an inverse proportion between the value of $\Lambda^*$ and the area $A_{xy}$. We propose now to apply a similar idea of producing subexperiments from the data to test this prediction, but instead of taking the full spatial information (full camera frame of view) and dividing the temporal data into subexperiments (as done in Fig.~\ref{fig:autocorr_ensemble}), here we do the opposite, we consider the full temporal information but split the spatial data into smaller regions $A_{xy}^{sub}$. The area of the full frame image seen in Fig.~\ref{fig:AV_fluctuation} was $840$x$840$ pixels  and we have approximately $25$ pixels per millimeter which means that each pixel corresponds to a linear extent of $40$ micrometers and an area of 1.6 \: 10$^{-3}$ mm$^2$. The total area of the $840$x$840$ pixels window used in the analysis up to this point is then $A_{xy}^{tot}$ = 1129 mm$^2$.

\begin{figure}[t]
	\centering
	\includegraphics[width=0.5\linewidth]{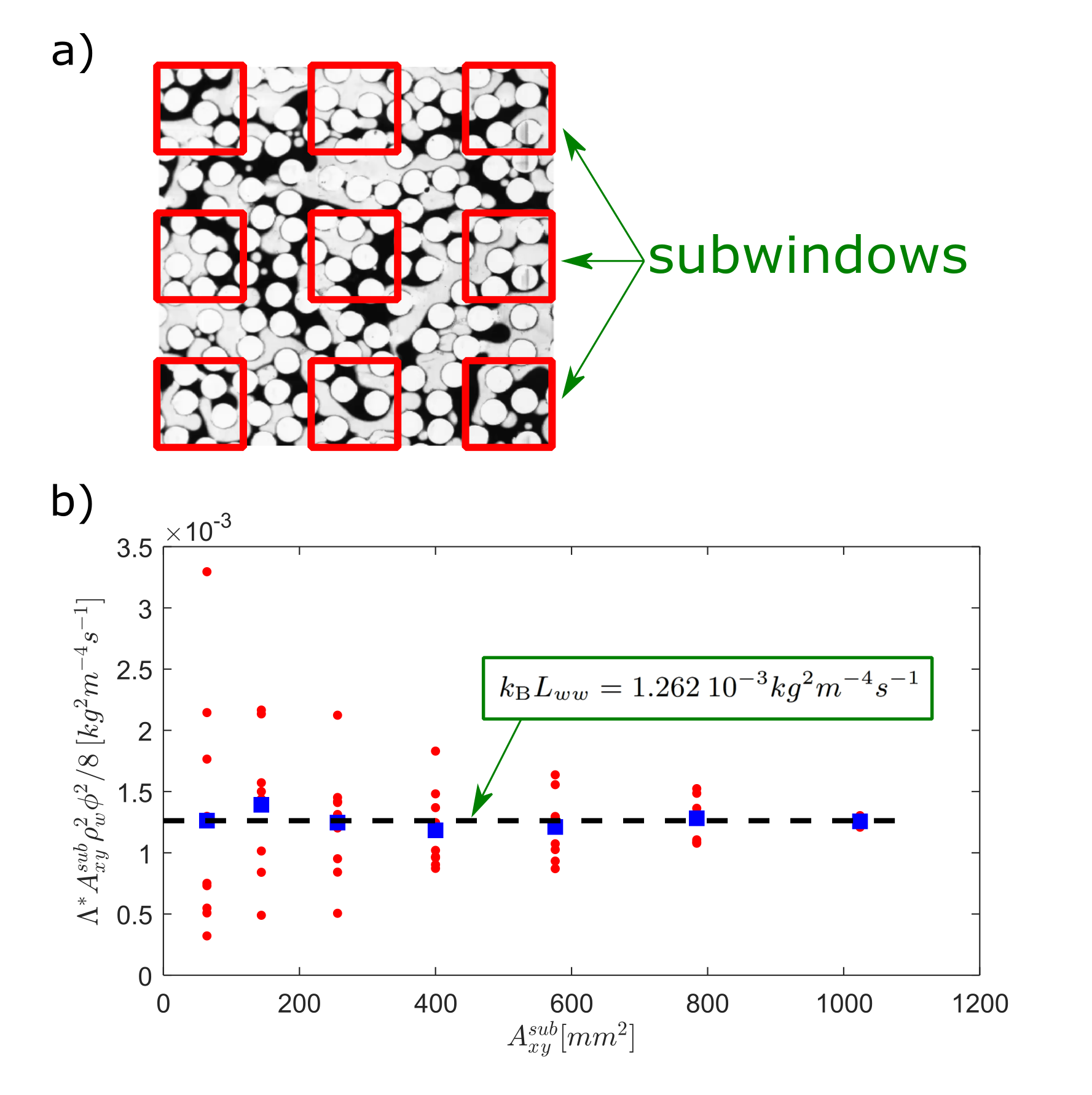}
	\caption{a) Splitting of the observation window into a set of $3$x$3$ subwindows. Each subwindow of area $A_{xy}^{sub}$ defines a new subexperiment for which we compute $\Lambda^*$. b) Plot of the reduced variable $\Lambda^* A_{xy}^{sub} \rho_w^2 \phi^2/8$ as a function of the subwindow area $A_{xy}^{sub}$. Each of the red dots correspond to the result of one subexperiment (i.e., the analysis performed using the whole time series observing the dynamics in a subwindow $A_{xy}^{sub}$. The blue squares show the average for each subwindow size. As expected from Eq.\ref{eq:lambdaomega} this quantity is indeed a constant, approaching the value marked by the dashed line which gives the measurement $\kappa_B L_{w w} = 1.262 \: 10^{-3}$ kg$^2$m$^{-4}$s$^{-1}$. }
	\label{fig:small_rev}
\end{figure}

For simplicity, we will split the initial area into square boxes of lateral extent in pixels $(800, 700, ...,300,  200)$. For each of those 7 different sizes, we will consider 9 boxes in a $3$x$3$ grid filling the whole initial $840$x$840$ frame of view, see Fig.~\ref{fig:small_rev}a). Notice that this means that for all except the smallest $200$x$200$ case (shown as red boxes in Fig.~\ref{fig:small_rev}a)), there is some degree of overlap between neighboring observation windows. We then perform the full analysis for each of the 63 subexperiments (7 different sizes, 9 subexperiments per size) to compute the asymptotic value $\Lambda^*$ of the integral of the autocorrelation function (as done previously for the whole image in Fig.~\ref{fig:autocorr}). If the relationship in Eq.~\ref{eq:lambdaomega} is correct, then we expect the reduced variable $\Lambda^* A_{xy}^{sub} \rho_w^2 \phi^2/8$ to be a constant equal to $\kappa_B L_{w w}$ and therefore independent of the area $A_{xy}^{sub}$. This prediction is indeed verified as can be seen in the plot in Fig.~\ref{fig:small_rev}b). The red dots in this figure correspond to the 63 subexperiments analyzed and the larger blue squares are the averages of each group of 9 subexperiments with the same area $A_{xy}^{sub}$. Notice that, although some statistical fluctuation is present, when considering the averages we see that indeed the plotted quantity is constant, from which, using Eq.~\ref{eq:lambdaomega}, we reach the result

\begin{equation}
    \kappa_B L_{w w} = 1.262 \: 10^{-3} \text{kg}^2 \text{m}^{-4}\text{s}^{-1} \:.
    \label{eq:kb_Lww}
\end{equation}

This result provides us with the product of the wanted coefficient $L_{ww}$ and a constant which plays the role of Boltzmann's $\kappa_B$, but is most probably not numerically equal to this constant. We do as of yet not know if such a constant applies to FDT of porous media. It was discussed by Alfazazi et al. \cite{Alfazazi2024} that the use of a time scale for sampling which is different from the molecular scale, may lead to deviations from $\kappa_B$, like what is known for granular materials \cite{edwards1989}.  The observation, that the fluctuations in good approximation are Gaussian in this context, is interesting. Alfazazi et al. did not find Gaussian distributions, however, so this may not affect the validty of the FDT.  This specific point deserves more attention in future works.

More work is also needed in order to find the volume permeability and the relative permeabilities of the component flows from their fluctuations. This is a topic of ongoing work and we present some preliminary ideas in the Appendix. The present procedure does not measure volume fluctuations. We are using the assumption of incompressible flows in such a way that the time rate of change of the saturation of one phase equals the rate of change of the saturation of the other phase (with opposite sign).

The present results contribute to the development of a coarse grained description of incompressible multiphase flow in porous media \cite{Kjelstrup2018,Kjelstrup2019,bedeaux2021,Bedeaux2024}. The basic assumptions proposed in these theoretical works for the construction of the REV, its size and fluctuations, conform with the present experimental results reported here.

\section{Conclusions}
\label{sec:conclusions}
In this work we have shown how the correlations in the fluid phase saturations can be used to extract information on the Onsager coefficients of porous media two-phase flows. We have measured the autocorrelation function $C_{ww}(\Delta t)$ of the time derivative of the wetting phase saturation and its integral $\Lambda(\Delta t)$ and showed how the asymptotic value $\Lambda^{\infty}$ of this integral could be estimated (in spite of the experimental noise) by computing the value $\Lambda^{*}$ of the first peak in $\Lambda(\Delta t)$. This value is linked to the Onsager coefficients via Eq.~\ref{eq:lambdaomega}. In particular, this equation predicts an inverse relationship between the value of $\Lambda^{*}$ and the area $A_{xy}$ of the observation window used in the analysis. This result is verified experimentally in Fig.~\ref{fig:small_rev} where the analysis was performed using subwindows of the whole frame of view with varying area $A_{xy}^{sub}$.

This work provides support for a procedure to select a REV of a porous medium \cite{Kjelstrup2018, Kjelstrup2019, bedeaux2021, Bedeaux2024}, with the purpose of finding a continuum description for multiphase flow in porous media. 
The basic assumptions used in the construction of the REV, its size and fluctuations, conform with the experimental results. In particular, the results shown in Figs.~\ref{fig:autocorr_ensemble} and \ref{fig:small_rev} demonstrate how a poorly chosen REV can lead to wide deviations in the measured quantities.

Two separate ways of computing $\Lambda^{*}$ were explored, one by considering a temporal average of the full experimental time series and another by producing a pseudo-ensemble of shorter subexperiments, performing the analysis for each of those subexperiments and then taking a pseudo-ensemble average. Both methods produced nearly indistinguishable results, an interesting finding although not entirely surprising as the subexperiments are ultimately derived from the same data set. Although the first technique (time averageing the whole data signal) is relatively easier to implement, the pseudo-averaging technique was more instructive in the sense that it allowed us to see how the plateau level in $\Lambda(\Delta t)$ is formed when averaging many subexperiments and how the level of this plateau $\Lambda^{\infty}$ is well estimated by the position of the first peak $\Lambda^{*}$ in the signal. By using our first technique, we have further investigated how the total time span of the dataset $t_{max}$ affects the stability of the measurement. As a rule of thumb, we found that an experiment needs to last at least 100 times the decorrelation time $\tau^*$ for the signal to produce a statistically significant (stable) measurement of the plateau level $\Lambda^*$, necessary in the calculation of the Onsager coefficient $L_{ww}$.

This work opens new avenues of research by investigating the informational content of the fluctuations in the phase saturation signal in porous media flows. The analysis employed can be translated to 3 dimensional datasets obtained for example via CT-imaging of real porous samples, although a number of technical challenges may arise. Most particularly, we have noticed that in order to get data of enough quality to produce the autocorrelation function and its integral (as seen in Figs. \ref{fig:autocorr} and \ref{fig:autocorr_ensemble}), the imaging system (here a machine vision camera) needs to acquire data that is both fast enough to allow the computation of derivatives and for a long enough interval to yield meaningful statistics. In many experimental scenarios, one may be forced to choose either one or the other extreme due to possible constraints related to memory and/or other experimental limitations. If we analyzed the data of only one of the subexperiments studied in Fig.~\ref{fig:autocorr_ensemble}, we could have made a very erroneous guess for the estimate of the first peak $\Lambda^{*}$ (the spread in the colored curves is about a factor of 10). This error grows for shorter total acquisition times (notice that the duration of each subexperiment $\Delta t^{sub} = 100 s$ in Fig.~\ref{fig:autocorr_ensemble} is already more than one order of magnitude larger than the estimated decorrelation time $\tau = 6.2 s$ and that would still be clearly not enough).

\section*{Acknowledgement}
The development of this work benefited from fruitful discussions with Knut Jørgen Måløy and Eirik Grude Flekkøy. MM, DB and SK authors are grateful to the Research Council of Norway through its Centers of Excellence funding scheme, project number 262644. MM additionally acknowledges the Research Council of Norway through its Researcher Project for Young Talent, project number 324555. R.T.A. acknowledges his Australian Research Council Future Fellowship FT210100165.

\bibliographystyle{ieeetr}
\bibliography{bib}

\newpage

\section*{Appendix: Considerations of flux-force relations}

In this appendix, we present some ideas that are necessary to establish a link between the classical relative permeability theory of two-phase porous media flows \cite{bear1972} and a description using non-equilibrium thermodynamics (NET). More details can be found in \cite{Kjelstrup2019, Bedeaux2024}. 

It is common to express the Darcy velocity of the separate fluid phases by:
\begin{equation}
u_w = V_wJ_w = -\frac{k_{rw}K}{\eta_w} \nabla p_w \:,
\label{eq:Jw32}
\end{equation}
\begin{equation}
u_n = V_nJ_n = -\frac{k_{rn}K}{\eta_n} \nabla p_n \:,
\label{eq:Jn32}
\end{equation}
where as usual $u_i$ is the Darcy velocity, $V_i$ is the partial molar volume in the REV (see definition later in Eq.~\ref{eq:Vi}), $k_{ri}$ denotes the relative permeability, $K$ is the absolute permeability of the porous medium, $\eta_i$ is the dynamic viscosity and $p_i$ is the pressure of the single bulk phase. In all quantities the subindex $i$ denotes the specific phase (wetting or non-wetting). The pressure gradients of the separate phases have been used.

The entropy production is composed of two flux-force products

\begin{equation}
    \sigma = - \frac{1}{T}\left (J_w \nabla \mu_w + J_n \nabla \mu_n  \right)
\end{equation}
Its value is invariant to change of variables. 
Here $J_w$ and $J_n$ are the fluxes of wetting and non-wetting components, respectively, in mol/(m$^2$s), $T$ is the temperature, and $\mu_w, \mu_n$ are the component chemical potentials \footnote{the notation for the chemical potential should not be confused with the dynamic viscosity which is typically associated to the letter $\mu$. Here we use the symbol $\eta$ for this viscosity.}. The gradients in these are the driving forces. For all practical purposes we are working with isothermal systems. The REV is a small thermodynamic system with the temperature  controlled by the environment, and we can measure its value in a normal way \cite{Bedeaux2024}.
It follows, that the linear flux-force relations can be written as
\begin{equation}
J_w = -\frac{L_{ww}}{T}\nabla {\mu_w} -\frac{L_{wn}}{T} \nabla {\mu_n} \:,
\label{eq:Jw1}
\end{equation}
\begin{equation}
J_n = -\frac{L_{nw}}{T}\nabla {\mu_w} -\frac{L_{nn}}{T} \nabla {\mu_n} \:\:.
\label{eq:Jn1}
\end{equation}
In order to connect the two descriptions, 
We shall use as definition of the pressure of a single phase $i=w,n$ 
\begin{equation}
   \nabla p_i \equiv \frac{1}{V_i} \nabla {\mu_i}
\end{equation}
where ${\mu_i}$ contains the effect of possible concentration variations as well as capillary pressures. The value of $\mu_i$ can be accessed by equilibrium conditions for the chemical potential at the REV boundary. The $V_i$ is the partial molar volume of $i$ or partial specific volume in the REV. It is defined for the REV by 
\begin{equation}
    V_i \equiv \left( \frac{dV^{\text{REV}}}{dN_i^{\text{REV}}} \right)_{T, N_j^{\text{REV}}}
    \label{eq:Vi}
\end{equation}
The value of $V_i$ can be computed when the REV composition is varied, $V^{\text{REV}}$ being the volume (here: area) of the REV and $N_i$ being the number of moles of phase i in the REV. For definitions of other thermodynamic variables of the REV we refer to \cite{bedeaux2021, Bedeaux2024}.

Under the assumption that the system is homogeneous, we can write

\begin{equation}
 V_i \equiv \left( \frac{dV^{\text{REV}}}{dN_i^{\text{REV}}} \right)_{T, N_j^{\text{REV}}} \approx \frac{V^{\text{REV}}}{N_i^{\text{REV}}} = \frac{V_p}{\phi \rho_i \frac{v_i}{M_i}} = \frac{M_i}{\rho_i \phi S_i} \:,
 \label{eq:ViMi}
\end{equation}
with $v_i$ being the volume in the REV occupied by phase $i$, $V_p = \phi V^{\text{REV}}$ the pore volume in the REV, $M_i$ the molar mass of phase $i$ and $S_i = v_i/V_p$ the saturation of phase $i$.

The resulting fluxes in Eqs.~\ref{eq:Jw1} and \ref{eq:Jn1} can then be written as:
\begin{equation}
J_w = -\frac{L_{ww}}{T}V_w\nabla {p_w} -\frac{L_{wn}}{T} V_n\nabla {p_n} \:,
\label{eq:Jw22}
\end{equation}
\begin{equation}
J_n = -\frac{L_{nw}}{T}V_w\nabla {p_w} -\frac{L_{nn}}{T} V_n \nabla {p_n} \:\:.
\label{eq:Jn22}
\end{equation}

Only independent, extensive variables can be used. An isothermal system of two  components has the entropy production $\sigma$. When the assumptions $\nabla p_w = \nabla p_n = \nabla p$ hold, we obtain:

\begin{equation}
    \sigma = J_V \left( - \frac{1}{T} \nabla  p \right) \:,
\end{equation}
where the volume flux is given by
\begin{equation}
J_V = J_wV_w + J_nV_n \:.
\end{equation}
The single flux-force relation becomes
\begin{equation}
J_V = -\frac{L_{VV}}{T} \nabla p 
\label{eq:JV3}
\end{equation}
The entropy production is invariant, i.e. it does not depend on the choice of variable set.

In the steady-state experiments considered here, the saturation is expected to be homogeneous across the sample (apart from regions close to the inlet and outlet boundaries where capillary end effects might occur). The capillary pressure $p_c = p_n - p_w$ is also expected to be spatially constant, which means that the assumption $\nabla p_w = \nabla p_n = \nabla p$ should hold. If we consider however a situation in which the pressure gradients are not necessarily the same, i.e., $\nabla p_w \neq \nabla p_n$, then by using Eqs.~\ref{eq:Jw22} and \ref{eq:Jn22} for $J_w$ and $J_n$ in Eqs.~\ref{eq:Jw32} and \ref{eq:Jn32}, we have

\begin{equation}
\left(\frac{k_{rw}K}{\mu_w}-\frac{V_wL_{ww}V_w}{T}\right)\nabla p_w -\frac{V_wL_{wn}V_n}{T}\nabla p_n = 0 \:,
\label{eq:sist1}
\end{equation}

\begin{equation}
-\frac{V_nL_{nw}V_w}{T}\nabla p_w +\left(\frac{k_{rn}K}{\mu_n}-\frac{V_nL_{nn}V_n}{T}\right)\nabla p_n = 0 \:.
\label{eq:sist2}
\end{equation}

This is a homogeneous system of equations for the pair $(\nabla p_w,\nabla p_n)$. For a non-trivial solution to exist, we need the determinant of the matrix of coefficients to be zero. In this simple 2x2 system this can also be seen by isolating $\nabla p_n$ from the second equation and substituting in the first one. After some algebra we are left with the following link between the two relative permeabilities and the Onsager coefficients:

\begin{equation}
\begin{aligned}
    \left(\frac{k_{rw}K}{\mu_w} - \frac{V_wL_{ww}V_w}{T}\right)
    \left(\frac{k_{rn}K}{\mu_n} - \frac{V_nL_{nn}V_n}{T}\right)
    = \left( \frac{V_wL_{wn}V_n}{T} \right)^2 \:.
\end{aligned}
\label{eq:sist_sol}
\end{equation}

Equation \ref{eq:sist_sol} provides a potential connection between the multiphase Darcy approach and the approach we have outline here based on non-equilibrium thermodynamics. If one of the relative permeabilities is known and the Onsager matrix is measured, the other relative permeability can be obtained via this equation. Interestingly, this equation seems to predict the correct qualitative dependence between the separate relative permeabilities: if $k_{rn}$ increases, $k_{rw}$ decreases and vice versa.

The formulation presented here is still in the early stages of development, and several aspects remain not fully understood. Specifically, the relationship between the thermodynamic properties of the REV, such as the chemical potential and its various contributions contained in Eq. 34, and the macroscopic Darcy framework is not yet entirely clear. The physical significance and the measurement methods for these quantities at scales relevant to porous media systems are still under discussion. We offer this formulation as a possible candidate for bridging non-equilibrium thermodynamics with the multiphase Darcy model, rather than as a definitive answer. Further research is required to consolidate the theoretical basis for this connection and to develop experimental protocols for accurately measuring the relevant quantities.

\end{document}